\documentclass[aps,showkeys,superscriptaddress]{revtex4-2}

\usepackage{graphicx}
\usepackage{amsmath}
\usepackage{amssymb}
\usepackage{floatrow,subfig}
\usepackage{xcolor}
\usepackage{soul}
\usepackage{hyperref}
\definecolor{mypurp}{RGB}{148,0,211}

\begin{document}

\title{Getting jammed in all directions: Dynamic shear jamming around a cylinder towed through a dense suspension}

\author{Olav R\o mcke}
\affiliation{Norwegian University of Science and Technology, Department of Energy and Process Engineering, NO-7491 Trondheim, Norway}
\author{Ivo R. Peters}
\affiliation{University of Southampton, Faculty of Engineering and Physical Sciences, Highfield, Southampton SO17 1BJ, UK}
\author{R. Jason Hearst}%
\email{jason.hearst@ntnu.no}
\affiliation{Norwegian University of Science and Technology, Department of Energy and Process Engineering, NO-7491 Trondheim, Norway}


\date{\today}

\begin{abstract}
Experimental results of towing a cylinder through a dense suspension of cornstarch and sucrose-water are presented. Focus is placed on the jamming fronts that exist in such systems. The literature has concentrated on the propagation of the jammed region under pushing, pulling or shearing conditions independently. How the different fronts interact and if the fronts are symmetric when generated simultaneously has remained unexplored. Investigating this is our main goal. With the current setup, we are able to view a continuous, quasi-2D field around the cylinder. As such, a new way of generating jamming fronts is presented whereby pushing, pulling and shearing can be examined simultaneously. In agreement with previous studies, the front propagates roughly twice as fast in the longitudinal direction compared to the transverse direction, which is attributed to a single underlying onset strain, regardless of orientation from the cylinder. Although the jamming front shows nearly perfect transverse symmetry, there is clear longitudinal asymmetry. This is evident in the velocity and strain fields, and is also detectable in the front propagation velocity and onset strain.

\end{abstract}


\maketitle


\noindent \textbf{NOTE:}\vspace{0.2cm}\\
This work is published as:\vspace{0.1cm}\\
O.\ R\o mcke, I.R. Peters, R.J. Hearst (2021) Getting jammed in all directions: Dynamic shear jamming around a cylinder towed through a dense suspension. \emph{Physical Review Fluids}, \textbf{6}:063301.\\ \url{https://doi.org/10.1103/PhysRevFluids.6.063301}\vspace{0.2cm}\\
The data for this study is freely available from the University of Southampton repository:\vspace{0.1cm}\\
\url{http://dx.doi.org/10.5258/SOTON/D1835}

\section{Introduction}

The transient development of a dynamically jammed region has been investigated extensively in experiments where the applied stress has been in the form of pulling \cite{Majumdar2017}, pushing \cite{Peters2014, Waitukaitis2012, Han2016, Han2019} or shearing \cite{Peters2016, Han2018, Han2019b}. In these works, the different cases have been addressed separately, but how these three scenarios interact in a single system has remained unexplored. By dragging a cylinder through a suspension of cornstarch and sucrose-water, we present a new way of producing dynamic jamming fronts where the system will have coexisting regions that experience pulling, pushing and shearing simultaneously in an observable fashion.

A granular suspension, consisting of solid particles suspended in a liquid, can exhibit both fluid- and solid-like behaviour. The observed behaviour depends primarily on the volume fraction and the applied stress, but is also influenced by factors such as the deformation history \cite{Han2016, Han2018, Wang2018, Pastore2011, Bi2011, Majmudar2005, Khandavalli2015, Fall2012}, particle surface chemistry \cite{James2018}, confinement effects \cite{Brown2012, Brown2014, Peters2013} and the shape of the particles \cite{Brown2011, James2019}. Above a critical volume fraction and at low stress, a granular suspension will be in a jammed state, identified by a finite yield stress \cite{Liu2010}. Jamming will occur at lower volume fractions under shear deformation when frictional contact between particles is present \cite{Bi2011}. Some suspension systems, such as cornstarch-water mixtures, can switch between frictional and frictionless states depending on the applied stress. This results in the ability to flow at low stress, but a dynamic transition into a jammed state at high stress, when the volume fraction is below, but close to, the critical volume fraction.

The assumption of smooth, force-free particles fall short of adequately describing this flow behaviour. This is particularly true for dense suspensions where particle-particle interactions may dominate the internal force structure of the suspension. With regards to cornstarch suspensions, it has been identified that friction \cite{Singh2018, Sivadasan2019, Tapia2019, Madraki2017, Fernandez2013} and repulsive forces between particles \cite{Brown2014, James2018, Guy2015}, play key roles when it comes to describing the extreme shear thickening that is typically observed. The steady state behaviour of such systems has been studied in both experiments \cite{ Brown2011, Pan2015, Hermes2016, SaintMichel2018, Han2018} and simulations \cite{Singh2018, Mari2015, Seto2013, Mari2014, Kawasaki2018, Heussinger2013, Guy2020, Morris2018}. The consensus appears to be that the observed thickening and jamming is the development of a frictional contact network between grains as the amount of applied stress increases. 

When a low stress is applied, the particles are not able to overcome the repulsive force, and the lubrication layer ensures that the suspension flows as if the particles were frictionless. Above a critical stress the particles come into frictional contact, resulting in an increase in resistance to flow. It is worth mentioning that potential hydrodynamic mechanisms have been proposed that do not require frictional contacts \cite{Jamali2019}. A framework for characterizing the flow of non-inertial, non-Brownian, dense suspensions, has been presented by Wyart and Cates \cite{Wyart2014}. Examples of the use of this model can be seen in both experimental \cite{Han2018, Hermes2016, Pan2015} and numerical \cite{Morris2018, Singh2018, Mari2015,Guy2020} studies. This model describes how continuous shear thickening (CST), discontinuous shear thickening (DST), and dynamic jamming is the result of a transition from lubricated (frictionless) interactions to frictional contacts between particles.

To explain the transient response of such systems, a few more concepts need to be introduced. How does the suspension transition from a liquid-like to a solid-like state, and how does this macroscopically propagate through the suspension? This leads us to the concept of dynamic jamming fronts, first introduced by Waitukaitis and Jaeger \cite{Waitukaitis2013}. This sparked the study of dynamic jamming fronts \cite{Majumdar2017, Peters2014, Han2016, Peters2016, Han2018, Waitukaitis2012}. The previously mentioned works indicate the existence of a onset strain \cite{Han2016, Han2018}. In dry granular systems, this is often presented as the material needing some deformation in order to build up a sufficient number of strong contacts between grains, that in turn increases the resistance to further deformation \cite{Wang2018, Pastore2011, Bi2011, Majmudar2005}. Large amplitude oscillatory shear (LAOS) experiments with cornstarch suspensions \cite{Khandavalli2015, Fall2012} further strengthen the evidence for an onset strain being an important parameter for the transient onset of dynamic jamming. The onset strain, which depends on how densely packed the suspension is, sets the ratio between the velocity if the jamming front and the perturbing body \cite{Peters2014, Peters2016, Han2016, Han2018, Waitukaitis2012}. An onset strain can also be used to explain why the jamming front propagates with different speeds relative to the direction of the perturbation~\cite{Han2016}.

The study of jamming fronts have up to this point been viewed in systems where the perturbation is singularly in the form of pulling, pushing or shearing. How the jamming front propagates around a towed body that experiences all three of these simultaneously remains unexplored. Here, we investigate the shape and the speed of the jamming front around a towed cylinder.  As indicated by the literature, we link the observed behaviour to an onset strain that accompanies the jamming front.


\section{Experiment}
The purpose of the experimental setup is to be able to generate, visualize and measure jamming fronts around a moving body. In this paper we present observations of how the jammed region propagates when a cylinder is towed through a suspension of cornstarch and sucrose-water solution. We begin with a description of the suspension itself in section~\ref{sec:Suspension}, followed by rheological measurements in section \ref{sec:Rheology} in order to identify a composition capable of dynamically jamming. This is followed by a description of the setup in section \ref{sec:Setup} and how the measurements were conducted in section~\ref{sec:Measurements}.

\subsection{\label{sec:Suspension}Suspension}
The suspension used in the present study is a mixture of cornstarch (\emph{maizena maisstivelse}) and sucrose-water. An important parameter for characterizing the suspension is its volume fraction ($\phi$) defined as the fraction of the volume occupied by solid; in section~\ref{sec:Rheology}, we will show that it is only in a specific range of packing fractions where dynamic jamming occurs. It is important to have a suspension that does not change characteristics over the course of an experiment. Particle settling is one such effect that is typically addressed by density matching the fluid phase to the solid phase \cite{SaintMichel2018, Pan2015}. Depending on the nature of the experiment, perfect density matching might not be necessary \cite{Hermes2016, James2018, Brown2012, Waitukaitis2012}. In order to increase settling time, we use a $50.0 \pm 0.3$\% by weight sucrose-water solution as the liquid phase with a density of $\rho_l=1230$~kg/m$^3$ and viscosity of $\eta_0=16$~mPa~s. Using this composition, we estimate settling to have a negligible effect over the time span of one experiment \cite{Garside1997, Richardson1997}. A cornstarch particle is porous \cite{Han2017}, so when mixing a sample, some time is needed for the starch to soak in the liquid. After a sample was homogeneously mixed, it was soaked and intermittently mixed for 1~hour.

With regard to the rheometer measurements, we expect the suspension at rest to settle roughly 6~$\mu$m in 1~minute for the highest volume fraction ($\phi=0.362$). As an up or down stress ramp takes 50~seconds, we do not expect to observe any consequence of settling at this time scale. For a dilute suspension, drift due to settling is higher. At the lowest volume fraction ($\phi=0.05$), we estimate 130~$\mu$m in 1~minute, However, for the rheometer data presented here, no significant drift in the data was observed during the course of one experiment.


\subsection{\label{sec:Rheology}Rheology}

Dense cornstarch suspensions are typically characterised as non-inertial, non-Brownian, frictional suspensions, and their rheological properties have been extensively investigated \cite{James2018, Brown2014, Brown2012, Pan2015, SaintMichel2018, Hermes2016,  Fall2012, Khandavalli2015}. The aim of this section is to verify that our suspension is able to dynamically jam and identify in what range of volume fractions ($\phi$) this will occur. 

By using the framework from Wyart and Cates \cite{Wyart2014} it is possible to identify the frictionless and frictional jamming volume fractions. In this model, the predicted viscosity is given by
\begin{equation}
    \eta \equiv \frac{\Sigma}{\dot{\gamma}}=\eta_0\left(1-\frac{\phi}{\phi_\text{eff}}\right)^{-2},
    \label{eq:WC1}
\end{equation}
where
\begin{equation}
    \phi_{\text{eff}}=f(\Sigma)\phi_m + [1-f(\Sigma)]\phi_0
    \label{eq:WC2}
\end{equation}
and
\begin{equation}
    f(\Sigma)=1-e^{-\Sigma/\Sigma^*},
    \label{eq:WC3}
\end{equation}
where $\Sigma$ is shear stress, $\eta$ is viscosity and $\phi_\text{eff}$ is the effective volume fraction at which viscosity diverges. The shear stress is assumed to be proportional to pressure $\Sigma=\mu P$, where $\mu$ is the effective friction coefficient. As a consequence, equations \eqref{eq:WC1}-\eqref{eq:WC3} can be written in terms of $\Sigma$ instead of $P$ \cite{Han2018}. For a detailed discussion, the reader is referred to the original paper \cite{Wyart2014}, or examples of how this theory has been applied and tested \cite{Guy2020, Han2018, Mari2015, Hermes2016}. What is important here, is that in the limits of low stress ($\Sigma \rightarrow 0$) and high stress ($\Sigma \rightarrow \infty$), the viscosity is quasi-Newtonian, diverging at $\phi=\phi_0$ and $\phi=\phi_m$, respectively \cite{Wyart2014}. In other words, when low stress is applied, the viscosity tends to infinity when approaching $\phi=\phi_0$. If a high stress is applied, the viscosity tends to infinity when approaching $\phi=\phi_m$. Evaluating~\eqref{eq:WC1} at the low and high stress limits results in
\begin{equation}
\label{eq:visc_low}
    \lim_{\Sigma \rightarrow 0}\eta=\eta_0\left(1-\frac{\phi}{\phi_0}\right)^{-2}
\end{equation}
and
\begin{equation}
\label{eq:visc_high}
    \lim_{\Sigma \rightarrow \infty}\eta=\eta_0\left(1-\frac{\phi}{\phi_m}\right)^{-2}.
\end{equation}
Following the method of \citet{Han2018}, the two critical volume fractions ($\phi_m$ and $\phi_0$) are determined by fitting these functions to the measured viscosities in their respective regimes. Most importantly: the range these two volume fractions span is the range of volume fractions where the suspension is able to dynamically jam. The characteristic stress ($\Sigma^*$) controls the stress at which the transition between the two regimes occurs \cite{Han2018, Wyart2014, Guy2015}, which we find by curve fitting our data to the Wyart and Cates model described above. As the Wyart and Cates model has singularities at $\phi\geq\phi_m$ for sufficiently high stress, some care needs to be taken in order to fit the data. Here we restrict ourselves to the data points used to find $\phi_m$ and $\phi_0$, similar to the method used by \cite{Baumgarten2019}, resulting in $\Sigma^*=10.2$~Pa.

An AR-G2 rheometer from TA-Instruments, with a rotating parallel plate geometry with a radius of $r=20$ mm was used. Cornstarch granules have been reported to have a diameter of about $d_p=15$~$\mu$m \cite{Han2018, Hermes2016, Majumdar2017}. A gap size larger than $h=0.7$ mm was used in order to ensure that a sufficient amount of particles occupy the space between the plates. The volume fraction for each sample is calculated as
\begin{equation}
\label{eq:pf}
    \phi=\frac{(1-\beta)m_s/\rho_s}{(1-\beta)m_s/\rho_s + m_l/\rho_l + \beta m_s/\rho_w},
\end{equation}
where $\beta$ is water content, $m_s$ is the starch mass, $\rho_s$ is the starch density, $m_l$ is the liquid mass, $\rho_l$ is the liquid density and $\rho_w$ is the density of water. The values of these parameters are listed in table~\ref{tab:param}. Note that equation~\eqref{eq:pf} does not account for porosity. This can be adjusted for by multiplying $\phi$ with $1/(1-\xi)$, where $\xi$ represents porosity. $\xi$ is expected to have a value of about $0.3$ in our setup \cite{Han2017}. Correcting for this would not change our conclusion. For every volume fraction, a sample was prepared according to equation \eqref{eq:pf}, which typically resulted in a measurement uncertainty in $\phi$ of $\pm0.005$. The sample was pre-sheared at the maximum stress for 60 seconds. Viscosity measurements were stress controlled, and was done by ramping up and down three times over roughly $50$~s (dense regime) up to $400$~s (diluted regime) for one ramp. Depending on the volume fraction, the stress ranged from 0.04 Pa to a maximum of $100$~Pa (high $\phi$), $400$~Pa (medium $\phi$) or 40 Pa (low $\phi$). The duration and ramp range was tuned to minimize surface deformations and drift. 

\begin{table}
\centering
\begin{tabular}{llll}
\multicolumn{1}{l|}{Parameter} & \multicolumn{1}{l|}{Value} & \multicolumn{1}{l|}{Uncertainty} & Unit      \\ \hline
\multicolumn{1}{l|}{$\rho_l$}  & \multicolumn{1}{l|}{1230}  & \multicolumn{1}{l|}{10}          & kg m$^{-3}$  \\
\multicolumn{1}{l|}{$\rho_w$}  & \multicolumn{1}{l|}{995}   & \multicolumn{1}{l|}{10}          & kg m$^{-3}$  \\
\multicolumn{1}{l|}{$\rho_s$}  & \multicolumn{1}{l|}{1630}  & \multicolumn{1}{l|}{20}          & kg m$^{-3}$  \\
\multicolumn{1}{l|}{$\beta$}   & \multicolumn{1}{l|}{11.2}  & \multicolumn{1}{l|}{0.2}         & \%      \\
\multicolumn{1}{l|}{$\eta_0$}  & \multicolumn{1}{l|}{0.016}    & \multicolumn{1}{l|}{0.001}    & Pa s 
\end{tabular}
\caption{Physical parameters for the suspension of cornstarch and sucrose-water solution.}
\label{tab:param}
\end{table}

\begin{figure}
    \sidesubfloat[]{\includegraphics[height=0.4\textwidth]{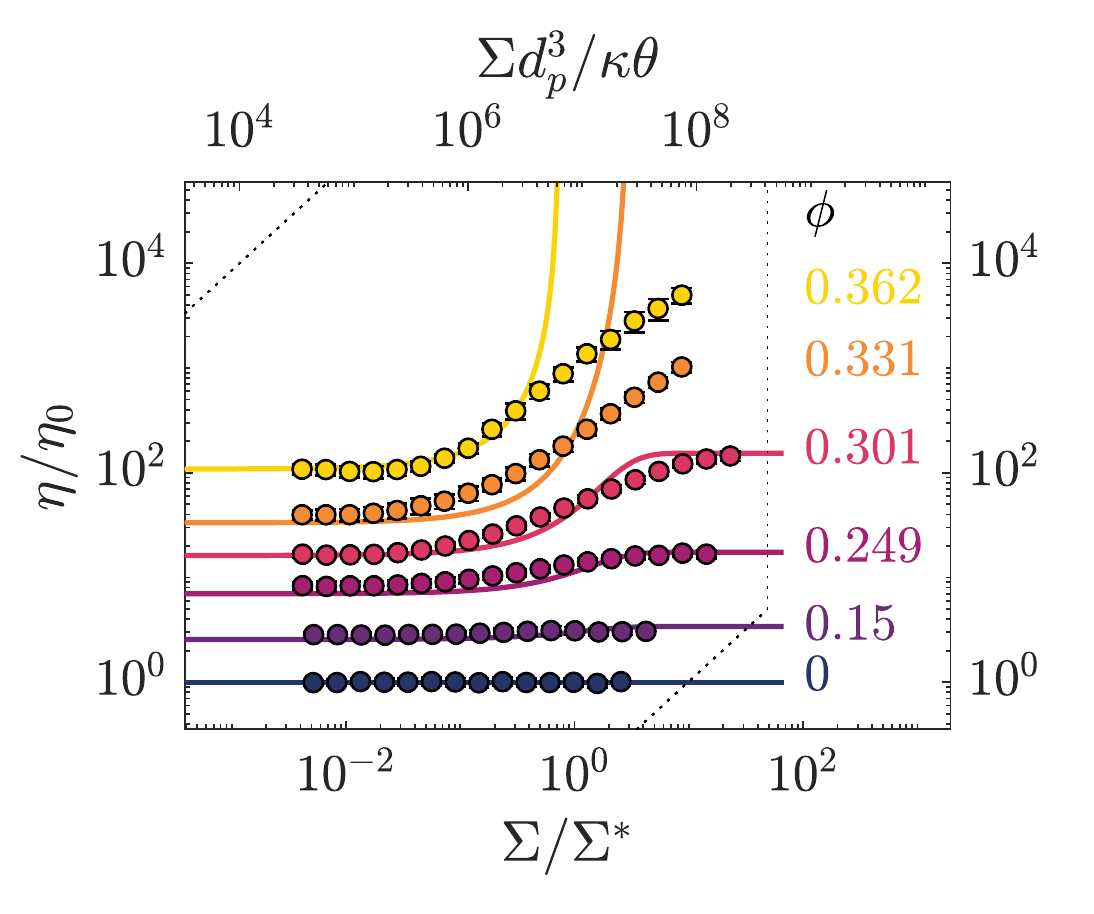}\label{fig:Exp:FlowCurve}}
    \sidesubfloat[]{\includegraphics[height=0.4\textwidth]{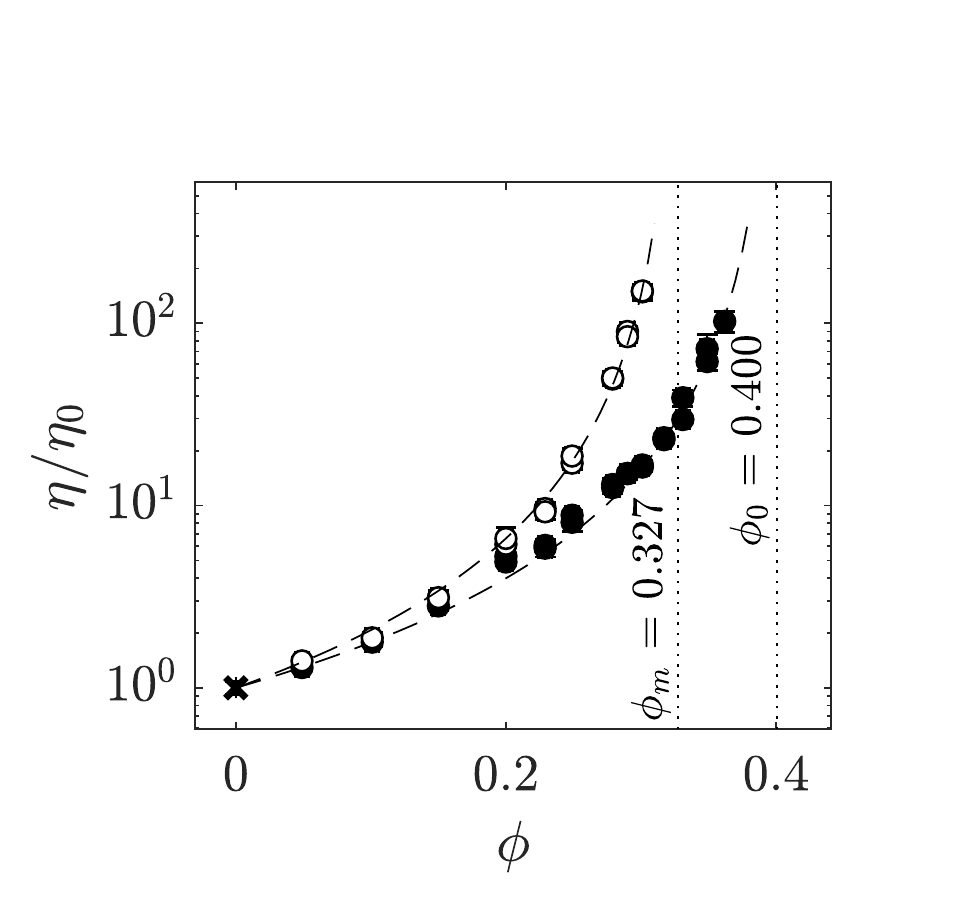}\label{fig:Exp:PackingFraction}}
   \caption{Rheology of the sucrose-water-cornstarch suspensions. (a)~Normalized viscosity as a function of applied stress. The volume fraction is indicated on the figure. The plotted lines are a result of a curve fit with the Wyart and Cates model described by equation \eqref{eq:WC1}-\eqref{eq:WC3}. We find $\Sigma^*$ to be $10.2$~Pa.  (b)~Normalized viscosity as a function of volume fraction in the frictionless low stress (filled), and frictional high stress (open) regimes. The dashed lines shows curve fits with equation \eqref{eq:visc_low} and equation \eqref{eq:visc_high}. Note that the two curves seems to diverge at different critical volume fractions $\phi_0$ and $\phi_m$ respectively, represented by the dotted lines.} 
   \label{fig:Exp:Rheo} 
\end{figure}

Finally, there are some limitations that restrict our measurement window. The rheometer itself can measure rotational speed up to $300$~rad/s with a resolution of $10^{-9}$~rad/s, and a maximum torque of $0.2$~Nm with a resolution of $10^{-9}$~Nm. This puts direct restrictions on stress and strain rate. In order to ensure non-Brownian, and non-inertial conditions, $Pe=\eta_0 d_p^3 \dot{\gamma} / \kappa \theta \gg 1$ and $St=\rho_s d_p^2 \dot{\gamma} / \eta_0 \ll 1$, respectively. Here, $\kappa$ represents Boltzmann's constant and $\theta$ is absolute temperature, kept at 20$^o$C for the entire experiment. The constraints in Peclet ($Pe$) and Stokes ($St$) number put additional restrictions on our measurement window. In addition, the surface tension should be sufficiently strong to hold the individual particles contained between the parallel plates. Here, we have used the semi-empirical relation $\Sigma_{max}\approx 0.1\Gamma/d_p$ \cite{Brown2012, Brown2014, Han2018, Peters2013}. $\Gamma$ represents the surface tension, which we set to 75~mN~m$^{-1}$ \cite{Schmidt2000}. For our system, this results in a maximum of roughly $500$ Pa. However, larger surface deformations were in some cases observed at lower stresses. These measurements were then discarded, and the maximum stress reduced. These restrictions reduced our measurement window, and are represented by the dotted lines in figure \ref{fig:Exp:FlowCurve}. 

\floatsetup[figure]{style=plain,subcapbesideposition=top}

Figure \ref{fig:Exp:FlowCurve} shows the viscosity response to applied stress for different volume fractions. Note the quasi-Newtonian regimes at high and low stress indicated by regions of constant viscosity for a change in applied stress. For $\phi> 0.3$ we are not able to reach the high stress Newtonian regime. In figure \ref{fig:Exp:PackingFraction} the viscosity at high and low stress have been plotted as a function of volume fraction. Note that at low stress, the viscosity seems to diverge at a volume fraction of $\phi_0=0.400$, while at high stress the viscosity diverges at $\phi_m=0.327$. If we consider a suspension with a volume fraction in-between these two values (vertical dashed lines in figure \ref{fig:Exp:PackingFraction}), the suspension flows, and behaves as a fluid, if we apply low stress to the sample, but the viscosity tends to infinity as stress is increased. In this study we choose a volume fraction of $\phi=0.36$.

\subsection{\label{sec:Setup}Towed cylinder experimental configuration}
\begin{figure}
    \centering
    \begin{minipage}{.5\textwidth}
        \sidesubfloat[]{\includegraphics[width=0.9\textwidth]{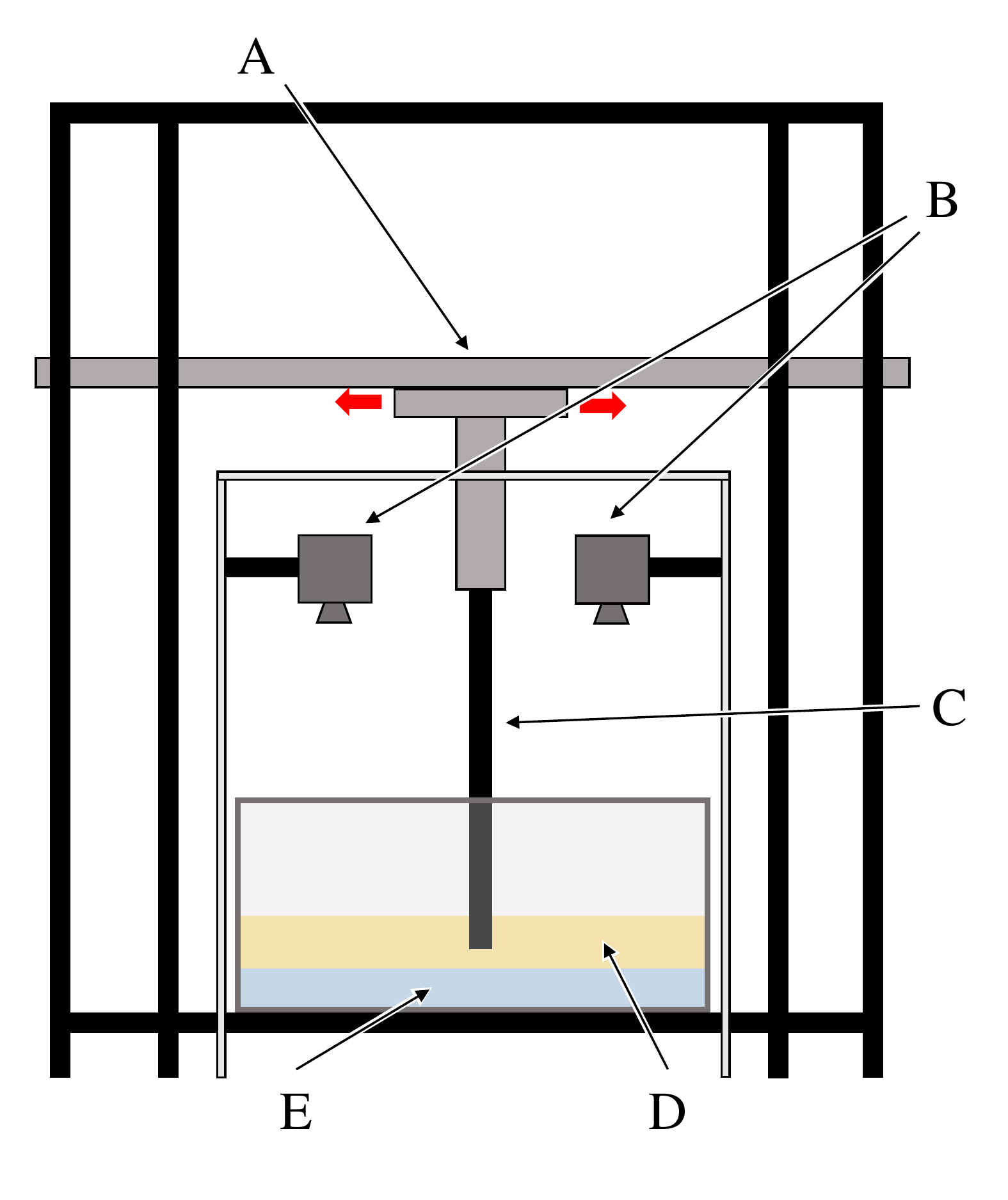}\label{fig:Exp:Setup}}
    \end{minipage}%
    \begin{minipage}{.5\textwidth}
        \begin{minipage}{1\textwidth}
             \sidesubfloat[]{\includegraphics[width=0.37\textwidth] {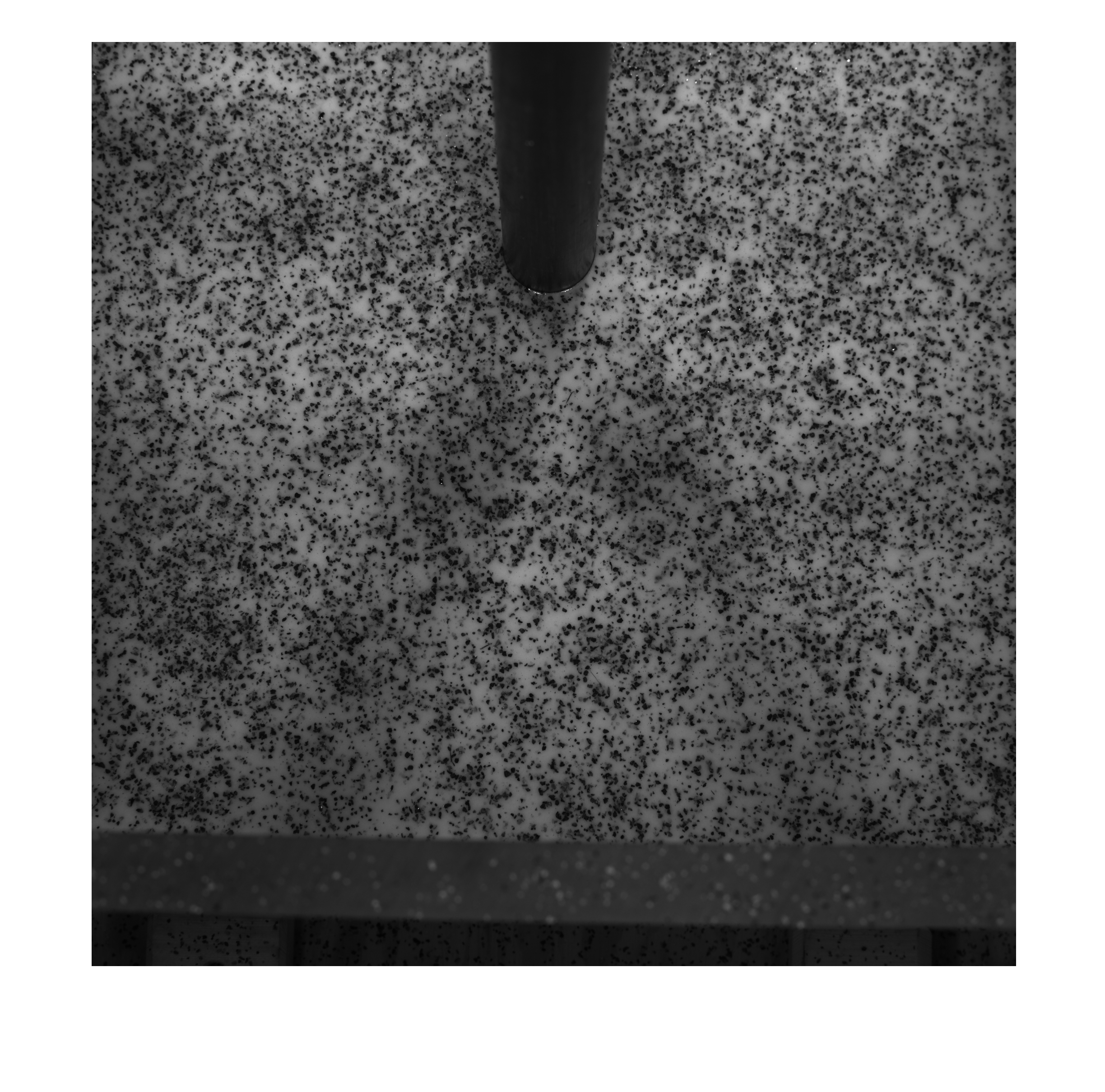}\label{fig:Exp:View}}
             \sidesubfloat[]{\includegraphics[trim= 1 1 10 10, clip,width=0.5\textwidth]{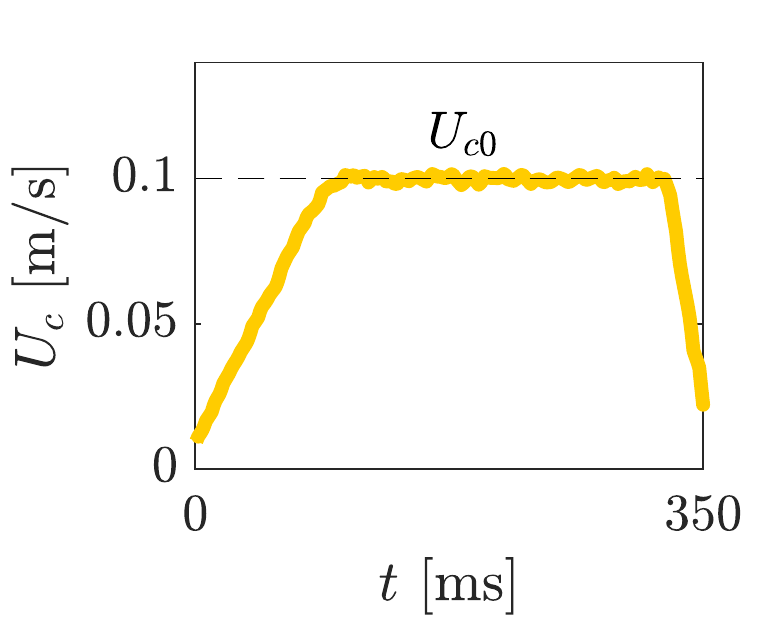}\label{fig:Exp:Trav}}
        \end{minipage}
        \vfill
    
         \sidesubfloat[]{\includegraphics[width=0.9\textwidth] {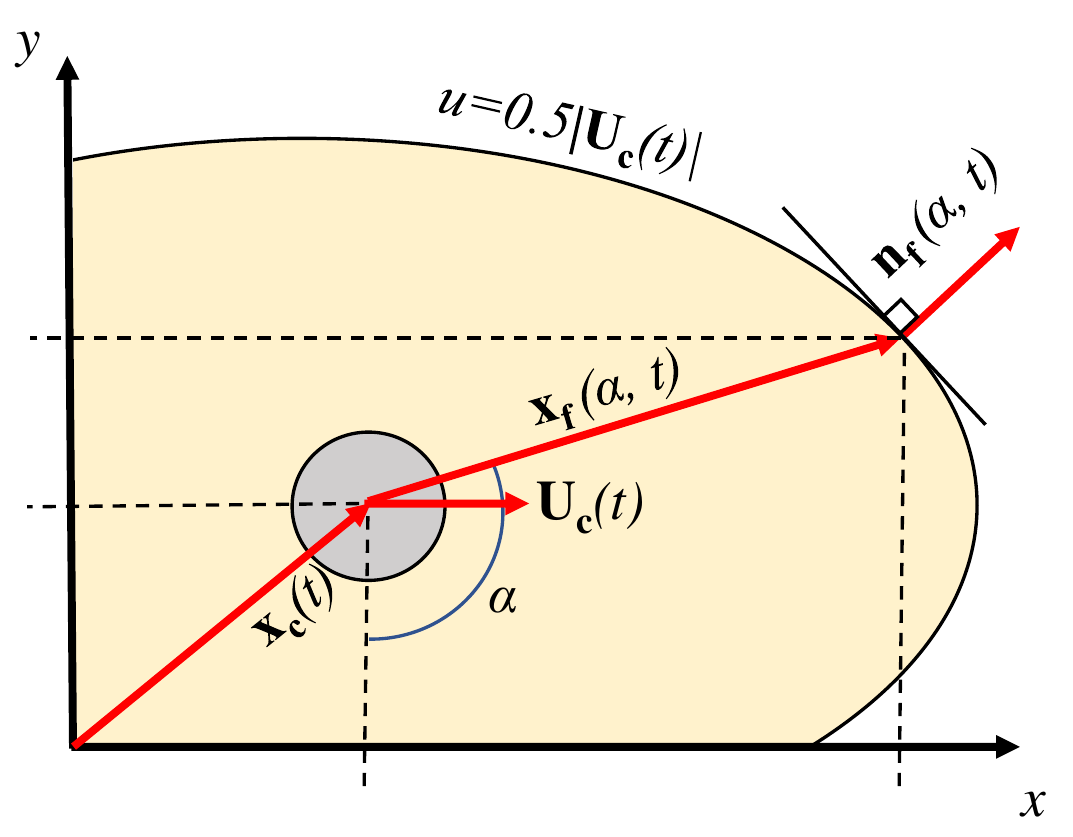}\label{fig:Exp:Def}}
    \end{minipage}
    \caption{(a) Schematic of the experimental setup with the traverse (A), cameras (B), cylinder (C), suspension (D) and Fluorinert (E). (b) Snapshot of the suspension surface. Black pepper is used as tracer particles for the PIV. See Supplemental Material \cite{SupMat} for a sample movie of both camera views. (c)~A time series of the cylinder velocity. Note that the traverse accelerates to constant velocity $U_{c0}$.  (d)~Schematic representation of the jamming front and variables where $\mathbf{x_c}$, $\mathbf{U_c}$, $\mathbf{x_f}$, $\mathbf{n_f}$ and $\alpha$, represents cylinder position, cylinder velocity, front position, front normal vector and angle, respectively. The jamming front is identified as the velocity contour $u=0.5|\mathbf{U_c}(t)|$. }
    \label{fig:Exp:Setup_Def}
\end{figure}

A schematic of the experimental setup is given in figure~\ref{fig:Exp:Setup}. The cylinder is driven by a Bosch-Rexroth MKR 15-65 traverse. Two Photron FASTCAM Mini WX100s, each with a resolution of 2048~pixels $\times$ 2048~pixels, are used for capturing the movement of the suspension surface. One camera is placed in front of the cylinder, and one behind.  The tank is 360~mm $\times$ 310~mm, while the cylinder has a diameter of $D=30$~mm. The suspension phase is $30$~mm thick, and floats on top of a 20~mm thick layer of high density fluorocarbon oil (Fluorinert, FC-74) which makes the system quasi two-dimensional \cite{Peters2014}. By comparing force and momentum change, \citet{Peters2014} showed that systems like this are close to 2D. Here, we vary the cylinder depth, and do an experiment at half the suspension layer thickness to confirm that there is no change in behaviour. The traverse drive is controlled through LabView, and sends a trigger signal for the cameras whenever the traverse starts moving. We drive the cylinder at 10 different velocities between $0.01$ and $0.14$~m/s. A frame rate is chosen such that the cylinder displacement between frames is roughly constant between cases. Black pepper serves as tracer particles for the PIV (Particle Image Velocimetry) analysis \cite{Peters2014, Peters2016, Han2018}. The pepper particle diameter is roughly in the range $3$ to $15$~pixels. In post-processing, the two resulting velocity fields are stitched together, making a continuous field around the cylinder. 

The following procedure was adopted prior to the actual measurements for each test case. First, a batch of cornstarch and water-sucrose solution was prepared, as explained in sections~\ref{sec:Suspension} and \ref{sec:Rheology}. The tank was then filled with Fluorinert, before the layer of suspension was poured on top. Finally, the cylinder was submerged $17$~mm into the suspension, before black pepper was sprinkled on the suspension surface. The view from one of the cameras is provided in figure~\ref{fig:Exp:View}. In figure~\ref{fig:Exp:View}, there is an average of 10 to 11 pepper grains per 48~pixel $\times$ 48~pixel square window, which is typical. After the measurements were conducted, the suspension layer was removed, and a new batch of suspension was poured onto the Fluorinert and the experiment repeated.

\floatsetup[figure]{style=plain,subcapbesideposition=top}

\subsection{\label{sec:Measurements}Measurement process}
Measurements were conducted by traversing the cylinder back and forth. This triggers the cameras, which capture the movement of the suspension surface until the camera memory was full. A time series of the cylinder velocity can be seen in figure \ref{fig:Exp:Trav}.

As mentioned, PIV was used to convert the resulting images into time-resolved velocity vector fields. In particular, LaVision Davis 8.4.0 was used. Three passes were done with square interrogation windows of 96~pixels $\times$ 96~pixels followed by two passes with circular interrogation windows with decreasing size for each pass, ending at 48~pixels $\times$ 48~pixels. This yielded one time-series of the velocity field in front, and one behind the cylinder. Each set of frames was stitched together with a weighted average in the overlapping region. If the cylinder blocked one of the camera views, the weight was set to zero in that region. The result was a velocity field where the only area not visible is the location where the cylinder penetrates the suspension surface. Finally, the resulting velocity field was time filtered with a moving Hanning window of 35 frames. Note that whether the cylinder is moving ``back" or ``forth" in the lab frame, the view is always rotated in post-processing, such that the cylinder is moving in the positive $x$-direction.

\begin{figure}
    \sidesubfloat[]{\includegraphics[trim= 0 5 0 10, clip, height=0.23\textwidth]{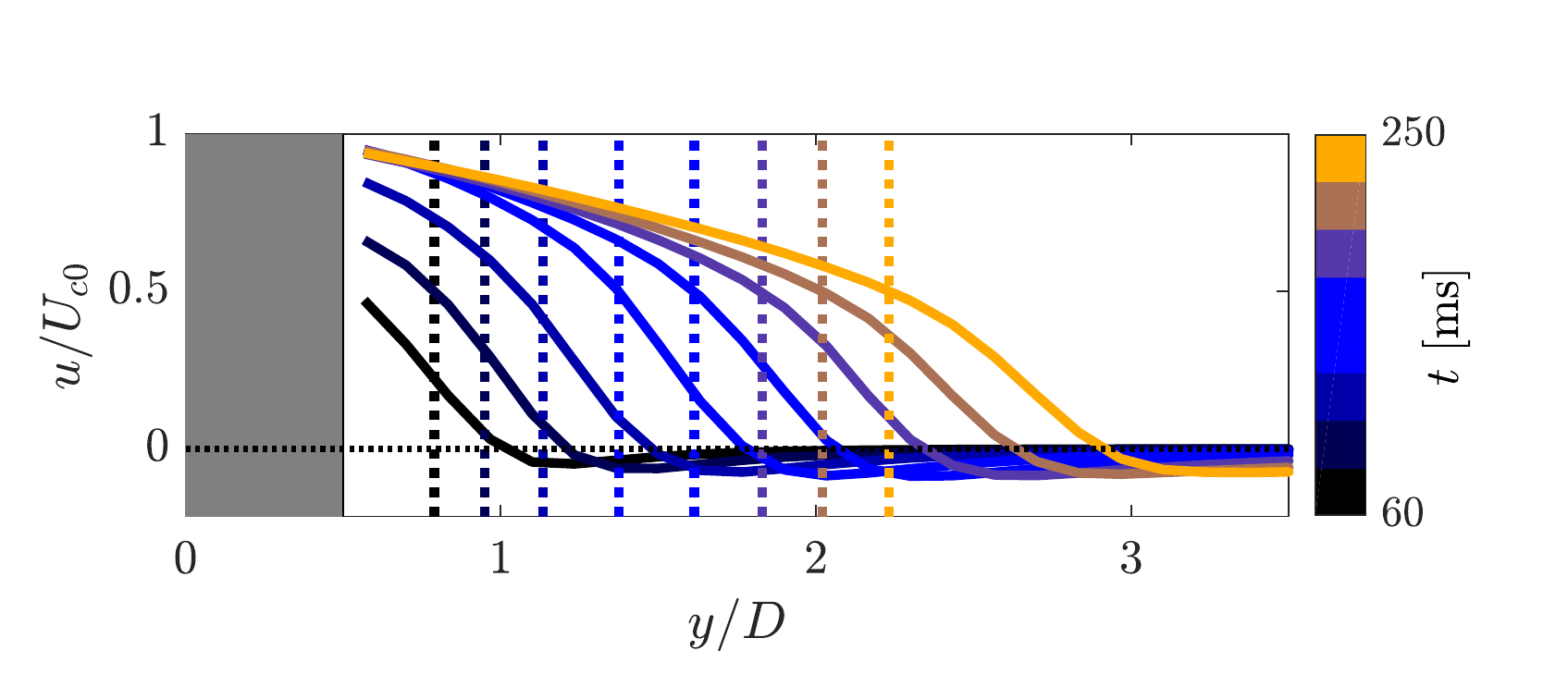}\label{fig:Exp:VelProf}}
    \sidesubfloat[]{\includegraphics[trim= 0 0 0 0, clip, height=0.23\textwidth] {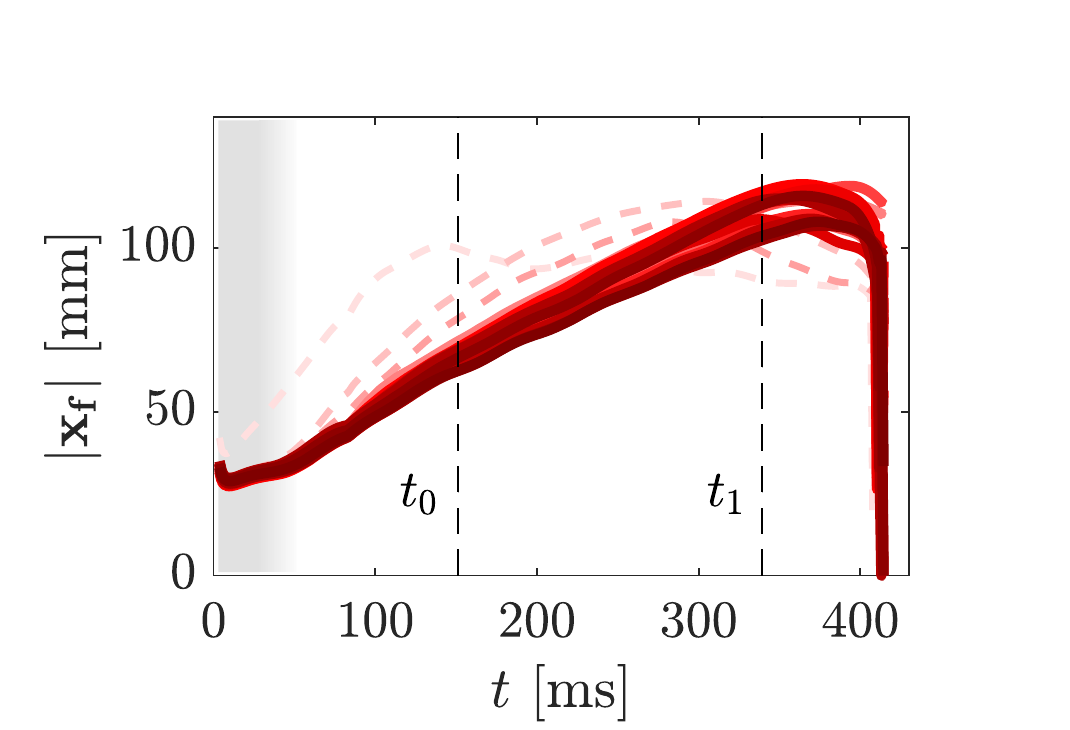}\label{fig:Exp:FrontPosTime}}
   \caption[]{The dynamic jamming front from two example experiments. (a) The velocity profiles at $\alpha=0$ and a cylinder velocity of $U_{c0}=0.14$~m/s. The black-blue-orange color scheme indicates time, and the gray region represents the cylinder. Dashed lines indicate the position of the jamming front with the corresponding color. Notice that the cylinder is accelerating for the first three profiles, but even though a constant cylinder speed is reached, the location of the front still propagates through the suspension. (b) The position of the jamming front at $\alpha=\pi/2$ and at a cylinder velocity of $U_{c0}=0.08$~m/s. The pink to red color scheme indicates consecutive measurements. The vertical dashed lines indicate the range where the cylinder has a constant speed, and the front is propagating freely through the suspension. The gray shaded region near $t=0$ represents the region where uncertainty is significant, and thus the results therein are not treated as statistically significant; they are included in this figure only for completeness.}
   \label{fig:Exp:JamFront} 
\end{figure}

The position of the jamming front is defined as the location in the suspension that has half the velocity of the cylinder \cite{Han2016, Han2018, Waitukaitis2012, Peters2014, Peters2016, Majumdar2017}. The first-order measures of the jamming front's location and orientation is indicated in figure~\ref{fig:Exp:Def}. Here, $\mathbf{x_c}$, $\mathbf{U_c}$, $\mathbf{x_f}$, $\mathbf{n_f}$ and $\alpha$, represents cylinder position, cylinder velocity, front position, front normal vector and angle, respectively. It is worth noting some key aspects of these  definitions. First, the front position ($\mathbf{x_f}$) is always relative to the cylinder. Furthermore, the cylinder is moving in the positive $x$-direction. Finally, $\alpha=0$ is orientated directly in the negative $y$-direction. As an illustrative example, figure \ref{fig:Exp:VelProf} shows the velocity profile of the suspension in the transverse cross-section ($\alpha=0$). The jamming front location is indicated by the dashed vertical lines. The front propagation factor $k_f$ will be studied in particular in section~\ref{sec:Front_Propagation}, and is defined as the speed of the front scaled with the speed of the cylinder. A more detailed description and derivation of the front propagation based on these front measures is given in appendix \ref{app:FrontProp}. 

\floatsetup[figure]{style=plain,subcapbesideposition=top}

Figure~\ref{fig:Exp:FrontPosTime} shows multiple time-series of the position of the jamming front in front of the cylinder. Note that from 0 to 150~ms, the cylinder is accelerating, and the jamming front is starting to develop. After 350~ms, the front interacts with the boundary. The region between the two dashed lines is the data used in the present analysis. In this region, the cylinder has a constant speed, and the front propagates through the suspension unimpeded. Note also that the first few runs show an earlier boundary interaction compared to the later runs. Through all cases, the first few runs differ from the later runs; the later runs are repeatable. Here, the analysis is done on the measurements in this repeatable regime, where the cylinder has a constant speed, and the front is propagating freely through the suspension. Both in dry granular materials~\cite{Bi2011} and dense frictional suspensions~\cite{Han2018} shear history is important. By excluding the first few runs, we, in effect, only consider experiments with the same deformation history.


\section{\label{Results} Results}

As previously mentioned, jamming fronts have been studied independently under pulling, pushing and shearing conditions~\cite{Han2016, Waitukaitis2012, Majumdar2017, Peters2014, Han2018, Peters2016}. The present setup allows for simultaneous observations of all three regimes. We begin by presenting the velocity field and the shape of the jamming front in section~\ref{sec:Velocity_Front}. This is followed by a description of the front propagation ($k_f$) and its relation to cylinder velocity ($U_{c0}$) and angle from the cylinder ($\alpha$) in section~\ref{sec:Front_Propagation}. To keep our variables in non-dimensional form, we will present the cylinder velocity scaled as $U_{c0}/U_c^*-1$ henceforth. Here, $U_c^*$ represents the velocity at which jamming fronts occur. At cylinder velocities lower than $U_c^*$, we cannot separate the front propagation factor ($k_f$) from zero. How we quantify $U_c^*$ is described in section~\ref{sec:Front_Propagation}. We identify that there is an onset strain~($\epsilon_c$) associated with the moving jamming front, not only in the longitudinal and transverse direction~\cite{Han2016}, but for all orientations about the cylinder. Further more, our setup makes it possible to directly compare the fore and aft half planes relative to the cylinder. We observe an asymmetry, which is also quantified.

\subsection{\label{sec:Velocity_Front}The velocity field and the jamming front}

\begin{figure}
    \sidesubfloat[]{\includegraphics[trim = 0 100 0 50, clip, width=0.97\textwidth]{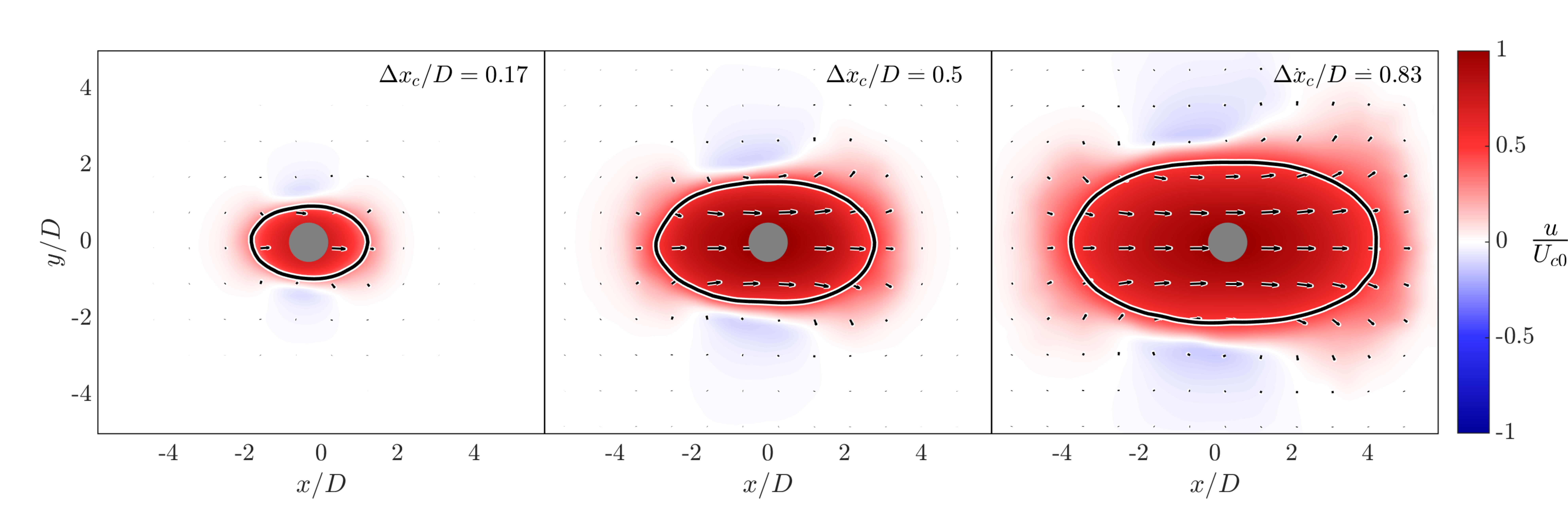}\label{fig:Res:VelocityField}}\\
    \sidesubfloat[]{\includegraphics[trim = 0 100 0 50, clip, width=0.97\textwidth]{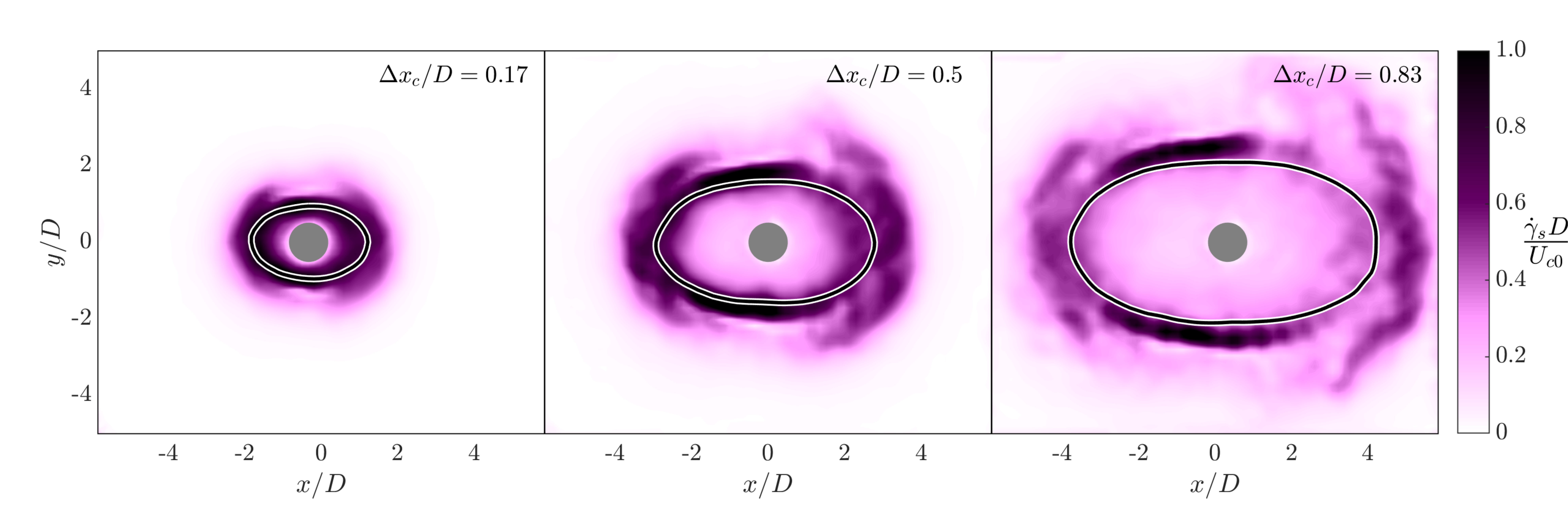}\label{fig:Res:RateField}}\\
    \sidesubfloat[]{\includegraphics[trim = 0 100 0 50, clip,width=0.97\textwidth]{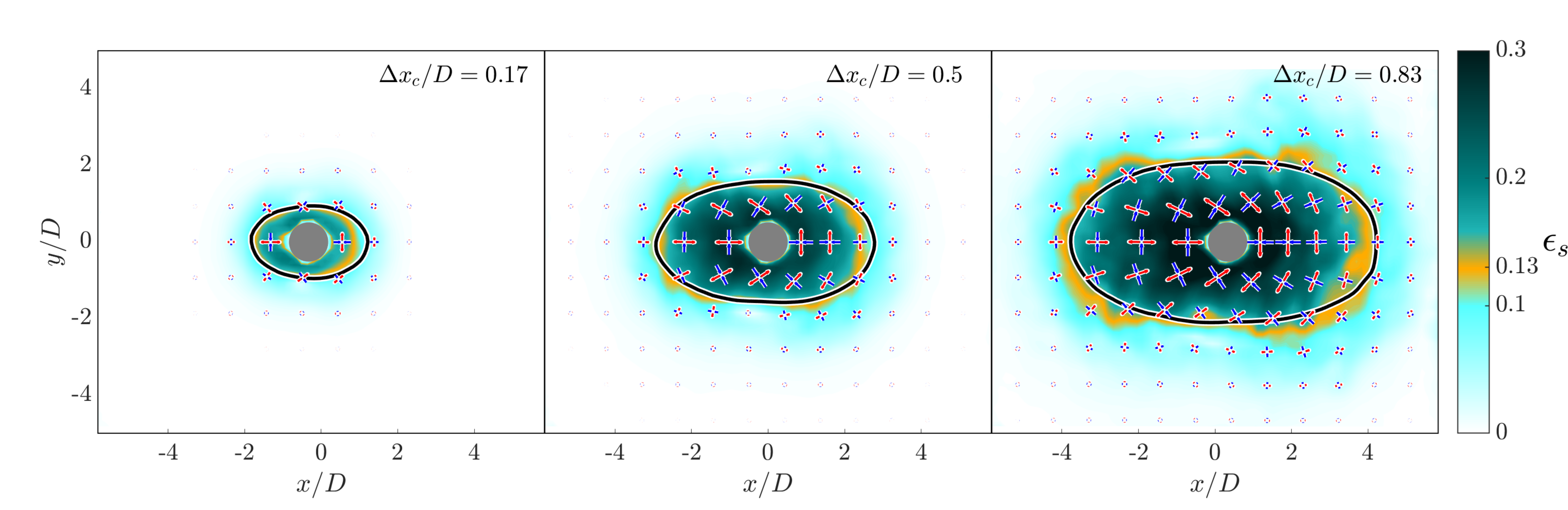}\label{fig:Res:StrainField}}\\
    \sidesubfloat[]{\includegraphics[trim = 0 0   0 50, clip, width=0.97\textwidth]{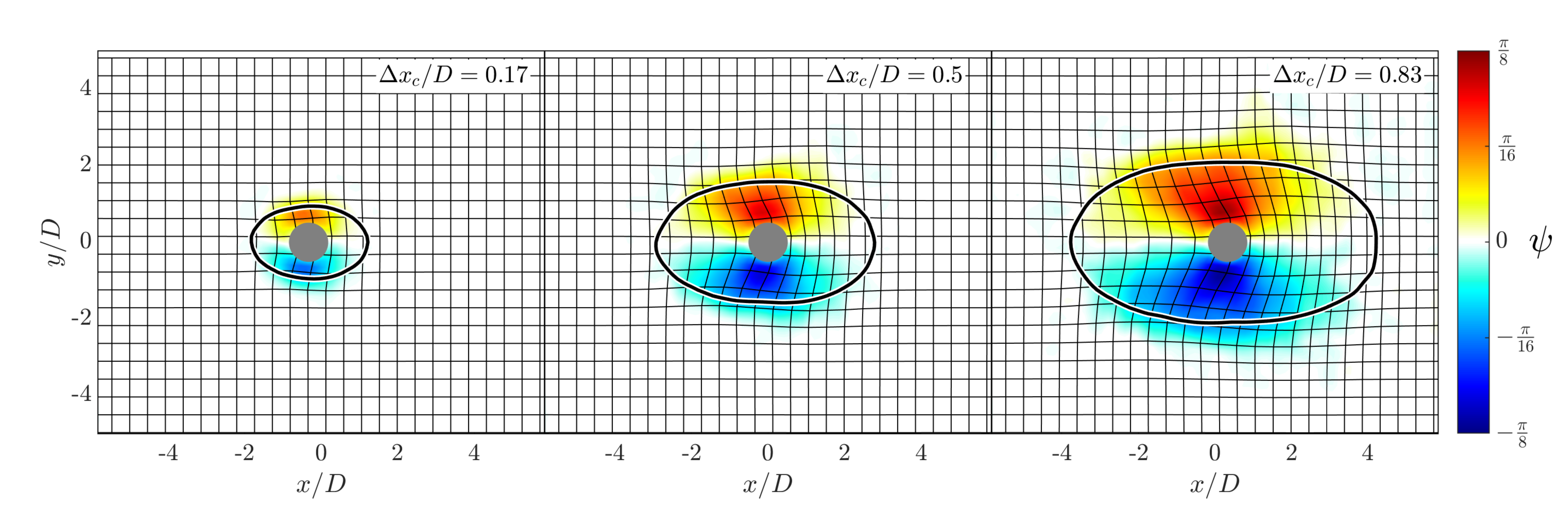}\label{fig:Res:RotField}}
   \caption[]{Snapshots of (a) the $x$-component of the velocity field with velocity vectors, (b) the deformation rate $\dot{\gamma}_s$ and (c) the accumulated strain $\epsilon_s$. The red arrows indicate direction of extension, while the blue arrows indicate the direction of compression. The length is scaled with $\epsilon_s$ to indicate amount of deformation. In addition, (d)~shows the rotation angle ($\psi$) of the material. The grid represents the movement of the material points. In all figures, the black line represents the jamming front and the cylinder velocity is at  $U_{c0}/U_c^*-1=5.4$. $\Delta x_c/D$ indicates the cylinder displacement normalised by the cylinder diameter. See the Supplemental Material~\cite{SupMat} for a sample movie of the four fields represented here.} 
   \label{fig:Res:Vel_rate_strain} 
\end{figure}

Figure~\ref{fig:Res:VelocityField} shows a time-series of contours of the $x$-component of the velocity field with velocity vectors superimposed. A region of near uniform velocity moves with the cylinder as it progresses through the suspension. This region of near uniform velocity grows over time, and is identified as the jammed region. The jammed region is separated from the unjammed region by the jamming front. The position of the jamming front ($\mathbf{x_f}$) is defined as the points in the suspension where the velocity is half the velocity of the cylinder \cite{Han2016, Waitukaitis2012, Peters2014, Peters2016, Majumdar2017}. A more thorough description is given in appendix \ref{app:FrontProp}. Note that the jamming front itself is symmetric both in the transverse, and longitudinal directions. However, in the unjammed region of the suspension, an asymmetry in longitudinal direction is observed. This can be seen by following the $u/U_{c0}=0$ contour.  

\floatsetup[figure]{style=plain,subcapbesideposition=top}

\begin{figure}
  \centering
  \includegraphics[trim = 0 0   0 50, clip, width=0.97\textwidth]{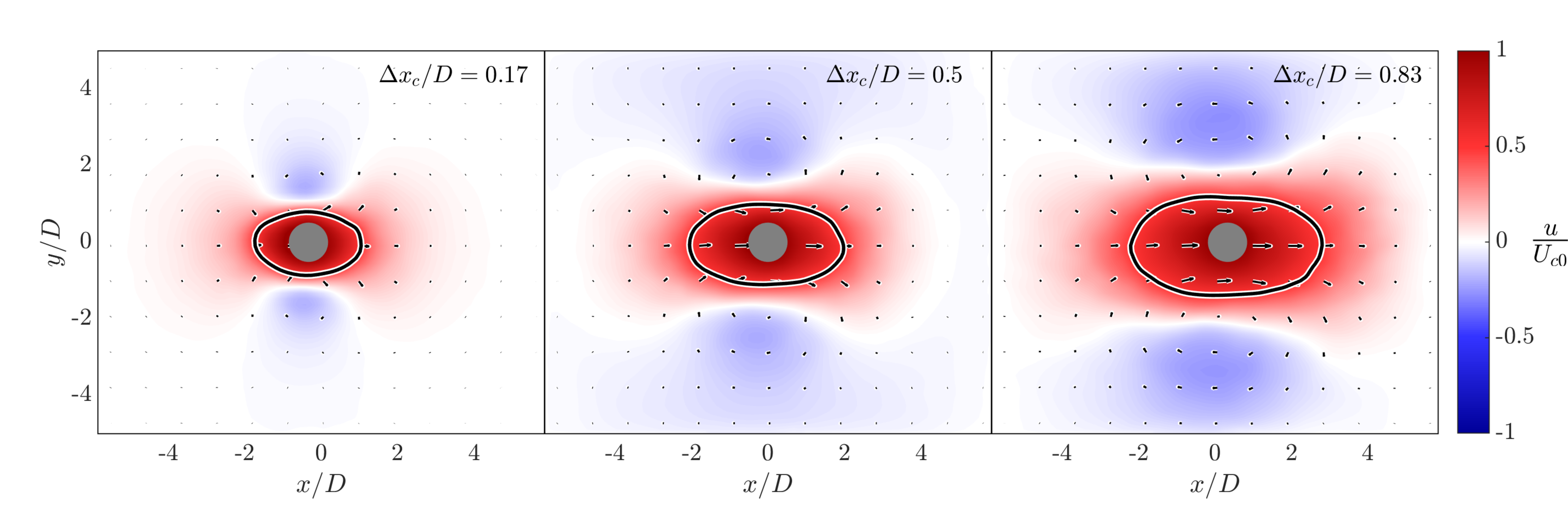}
  \caption{$x$-component of the velocity field at $U_{c0}/U_c^*-1=0.4$ with velocity vectors. The black line represents the jamming front. $\Delta x_c/D$ indicates the cylinder displacement normalised by the cylinder diameter. See the Supplemental Material~\cite{SupMat} for a sample movie of the field represented here.}
  \label{fig:Res:VelocityField_low}
\end{figure}

Figures \ref{fig:Res:RateField} and \ref{fig:Res:StrainField} represent the measures for strain rate $\dot{\gamma}_s$ and accumulated strain $\epsilon_s$. In short, we use the norm of the respective tensors to represent the intensity. The tensor for strain rate is the strain rate tensor, while the tensor for accumulated strain is the Eulerian logarithmic strain tensor~\cite{Nasser2004}. With regards to jamming fronts, earlier attempts at representing accumulated strain have typically been in the form of numerically integrating components of the strain rate tensor \cite{Han2016, Majumdar2017, Han2018}. We choose to represent strain by the Eulerian Logarithmic strain tensor, as it still is defined from the sum of increments, and represents deformation in the deformed configuration. The latter will be important, as we will compare the accumulated strain and the location of the jamming front. A description of how $\dot{\gamma}_s$ and $\epsilon_s$ are assessed can be found in appendix~\ref{app:Strain}. 

From figure \ref{fig:Res:RateField} we see that a wave of high strain rate leads the way for the jamming front. As the wave passes through, it leaves behind a region of jammed material. The longitudinal asymmetry in the unjammed region mentioned earlier is also clear from this figure. At the jamming front in figure~\ref{fig:Res:StrainField}, $\epsilon_s$ has an approximately constant value throughout the experiment. We have indicated this with a colourband. This is discussed further in later sections. The behaviour described above is the case for $U_{c0}/U_c^*-1\geqslant2.2$; a wave of high strain rate travels in front of the jamming front. Similar behaviour was observed by \citet{Peters2014}. This differs from dry granular study of jamming fronts, where the highest change in volume fraction almost perfectly aligns with the jamming front \cite{Waitukaitis2013}. 

For lower $U_{c0}$, the jammed region is not as obvious. Figure \ref{fig:Res:VelocityField_low} shows the velocity field for  $U_{c0}/U_c^*-1=0.4$. Relative to the cylinder displacement, the front does not propagate as fast through the suspension, and the close-to-uniform-velocity region is not as large as what we see in figure \ref{fig:Res:VelocityField}. Compared to the high velocity cases, we see a longitudinal symmetry in the unjammed region of the suspension.

\begin{figure}
  \centering
  \includegraphics[width=\textwidth]{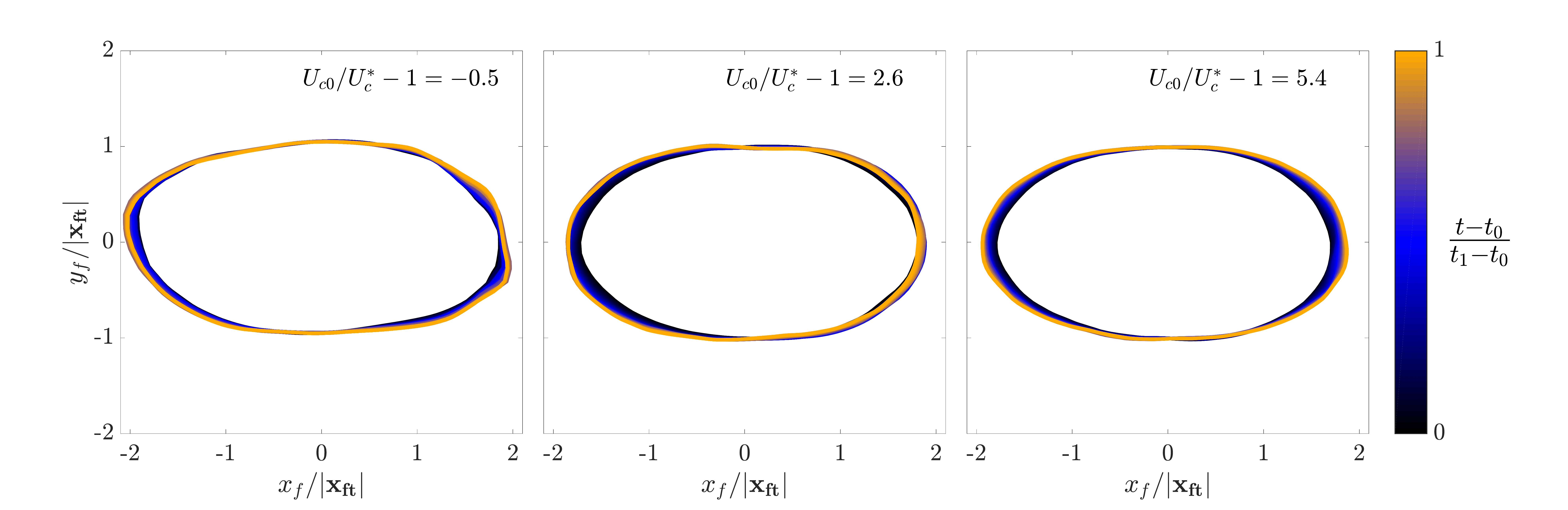}
  \caption{The shape of the $0.5U_c$ contour at three different velocities, indicated in the upper right in all figures. The color bar indicates time. $t_{0}$ is the time at which the cylinder has accelerated to a constant speed, while $t_{1}$ is when the cylinder stops, or the shape is influenced by interactions with the boundary. $t_0$ and $t_1$ are indicated by the dashed vertical lines in figure \ref{fig:Exp:FrontPosTime}.}
  \label{fig:Res:FrontPositionScaled}
\end{figure}

From the velocity field shown in figure \ref{fig:Res:VelocityField}, we see that for the duration of an experiment the area enclosed by the jamming front increases. Figure~\ref{fig:Res:FrontPositionScaled} shows the self-similarity of the shape of the jamming front. Here, the jamming front has been scaled by $|\mathbf{x_{ft}}|=\frac{1}{2}(|\mathbf{x_f}|_{\alpha=\pi} + |\mathbf{x_f}|_{\alpha=0})$. The shape of this region is approximately self-similar in the steady regime of the experiment. Note the anisotropic shape, where the longitudinal position of the front is close to a factor of 2 larger than the transverse position. Both the self-similarity, and the longitudinal-transverse anisotropy of the shape are consistent with earlier observations~\cite{Han2016, Peters2014}.

\subsection{\label{sec:Front_Propagation}Front propagation}

The front, $\mathbf{x_f}$, propagates through the suspension with a certain speed. Note that the jamming front is not associated with any specific material point, but a velocity contour. Thus, defining a propagation direction can be complex. Here, we define the direction of propagation as the normal vector of the jamming front. We will focus our analysis on the front propagation factor ($k_f$), which is the front propagation speed normalized by cylinder velocity, and is calculated as
\begin{equation}
\label{eq:front_norm_vel_norm}
    k_f = \frac{u_f}{|\mathbf{U_c}|} = \frac{u_f}{U_c}.
\end{equation}
Note that for a given $\alpha$ and $U_{c0}$, the data for $k_f$ presented here is averaged values, thus not a function of time. A more detailed description can be found in appendix \ref{app:FrontProp}.

Figure~\ref{fig:Res:FrontPropagationAlpha} shows the front propagation as a function of the angle $\alpha$ from the cylinder. Similar to figure~\ref{fig:Res:FrontPositionScaled} showing the scaled position of the front, the fact that the front propagation is different in the longitudinal and transverse directions is evident. In figure~\ref{fig:Res:FrontPropagationAlpha} this manifests as the two peaks in $k_f$ at $\alpha=\pi/2$ (in front of the cylinder) and $\alpha=3\pi/2$ (behind the cylinder), respectively. The inset of figure~\ref{fig:Res:FrontPropagationAlpha} demonstrates that if each $k_f$ curve is normalised by its peak value, that all tests with $U_{c0}/U_c^*-1 \ge 2$ fundamentally have the same shape. It is worth pointing out that the peak value is always at $\alpha=3\pi/2$, i.e., behind the cylinder, illustrating a repeatable asymmetry in the system.

\begin{figure}
  \centering
  \includegraphics[width=0.5\textwidth]{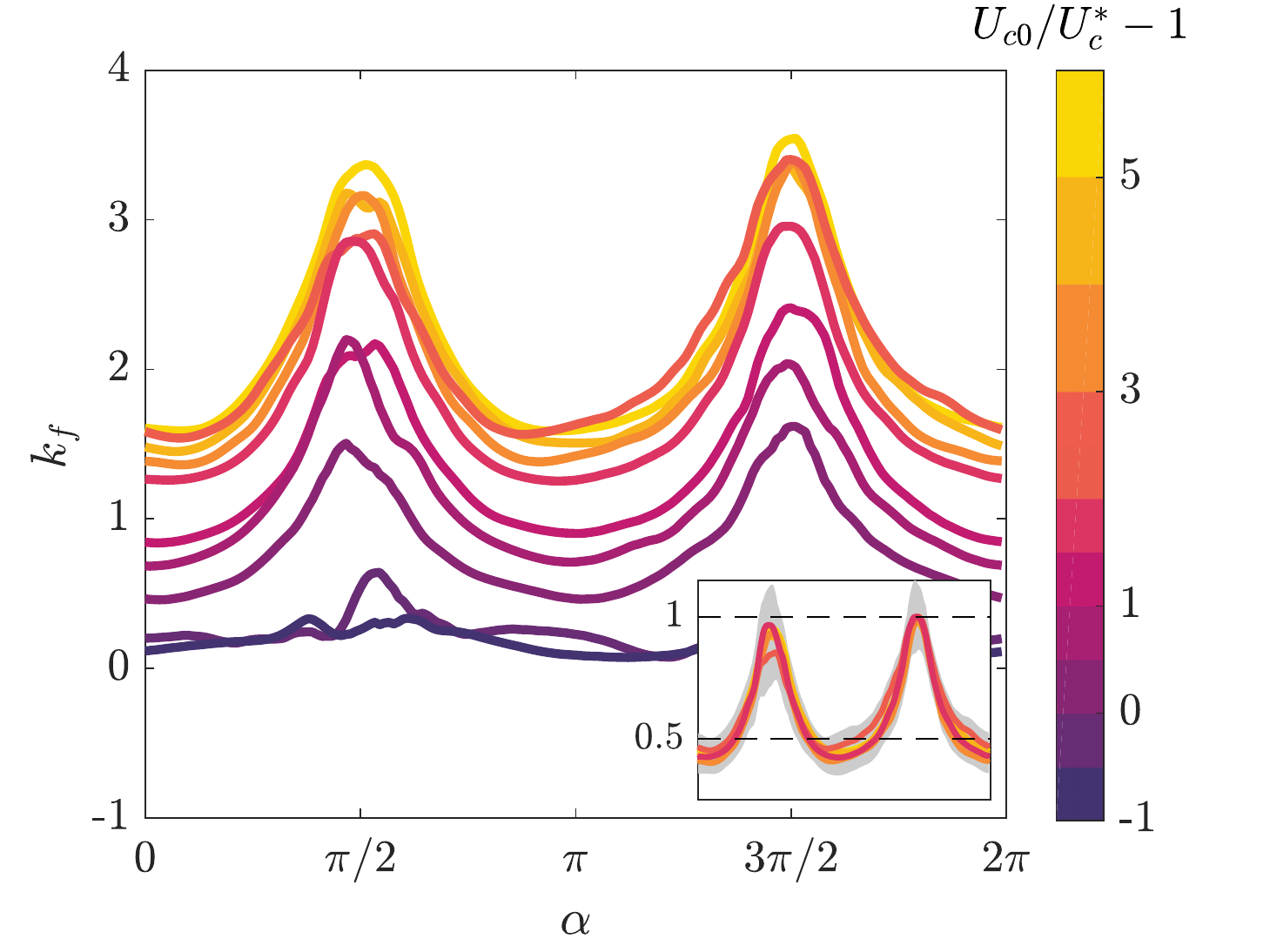}
  \caption{Front propagation factor as a function of the angle from the cylinder. The inset shows the same plot for $2 \le U_{c0}/U_c^*-1 \le 6$, but each curve from the main figure has been normalized with its maximum value. The gray region in the inset indicates one standard deviation from the collection of the time series. The peak at $\alpha=3\pi/2$ consistently has the highest value in these cases.}
  \label{fig:Res:FrontPropagationAlpha}
\end{figure}

Another aspect of interest is how $k_f$ develops as a function of the cylinder velocity $U_{c0}$. Below some non-zero cylinder velocity ($U^*_c$), $k_f\approx0$, meaning that the front is not propagating, thus the system is not jamming. On the other hand, as the cylinder velocity becomes sufficiently high, $k_f$ tends towards some constant value $k_f^*$. That is, the $k_f-\alpha$ curves in figure \ref{fig:Res:FrontPropagationAlpha} overlap for sufficiently high $U_{c0}$. The relationship between $k_f$ and $U_{c0}$ is investigated further by fitting the data to the curve \cite{Han2016}
\begin{equation}
\label{eq:front_norm_vel_curvfit}
k_f = \left\{ 
\begin{array}{rcl} 
    &0 & \mbox{for} \ \ \ U_{c0} \leq U^*_{c}, \\ 
    &k_f^*(1-e^{1-U_{c0}/U^*_{c}}) & \mbox{for} \ \ \ U_{c0} > U^*_{c}.
\end{array}
\right.
\end{equation}
We expand upon the work of \citet{Han2016} by also looking at intermediate angles about the cylinder. Equation~\eqref{eq:front_norm_vel_curvfit} has two free parameters; $U^*_{c}$ and $k_f^*$. The physical meaning of these two values can be interpreted as: $U^*_{c}$ represents the velocity below which jamming does not occur, and $k_f^*$ represents the value $k_f$ tends towards as $U_{c0}$ becomes sufficiently high. The curve fit is performed by finding the minimum rms between the measured data and the curve from equation~\eqref{eq:front_norm_vel_curvfit}.

\begin{figure}
    \sidesubfloat[]{\includegraphics[height=0.4\textwidth]{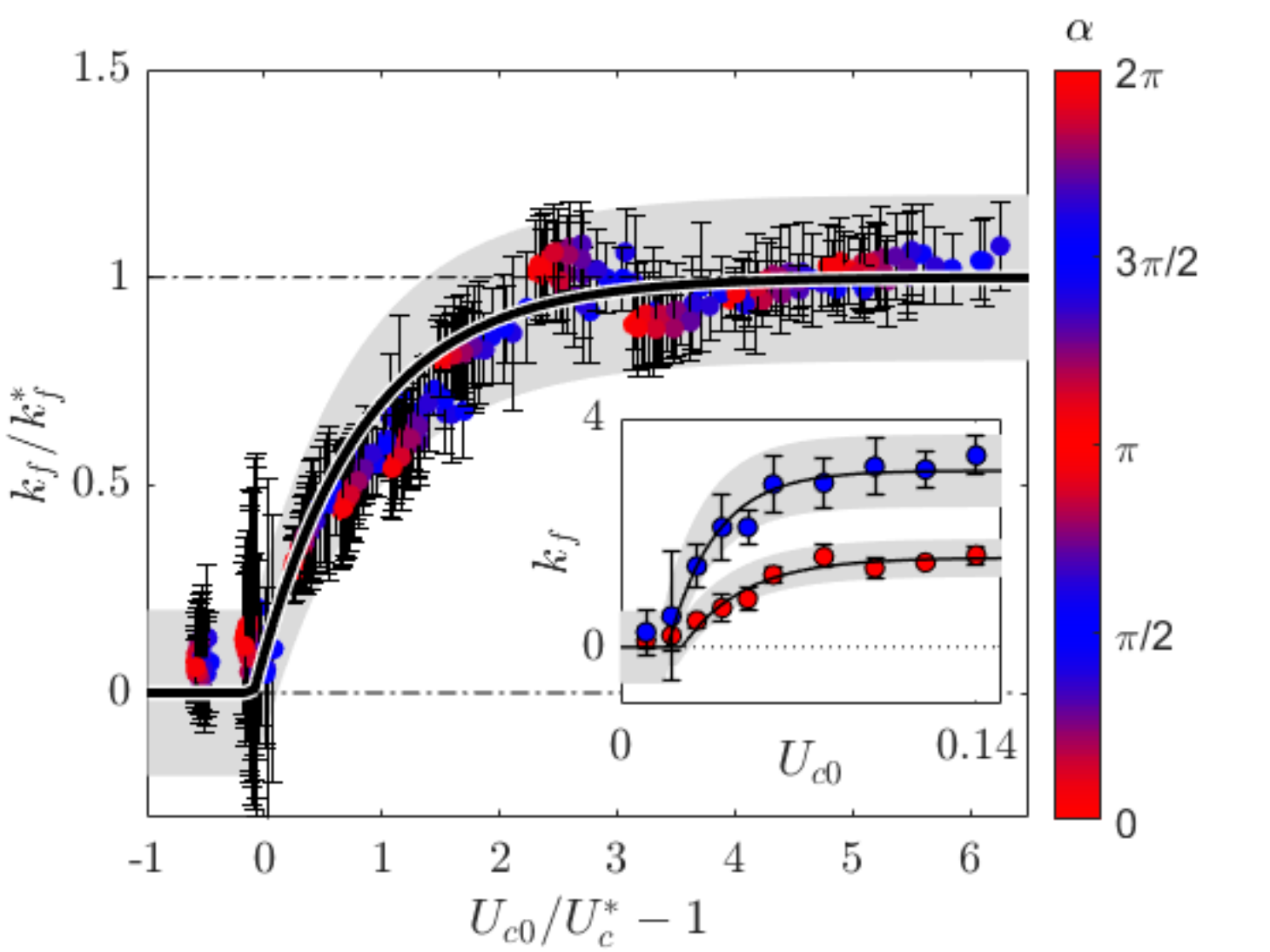}\label{fig:Res:NormalizedFrontPropagation}}\quad
    \sidesubfloat[]{\includegraphics[height=0.4\textwidth]{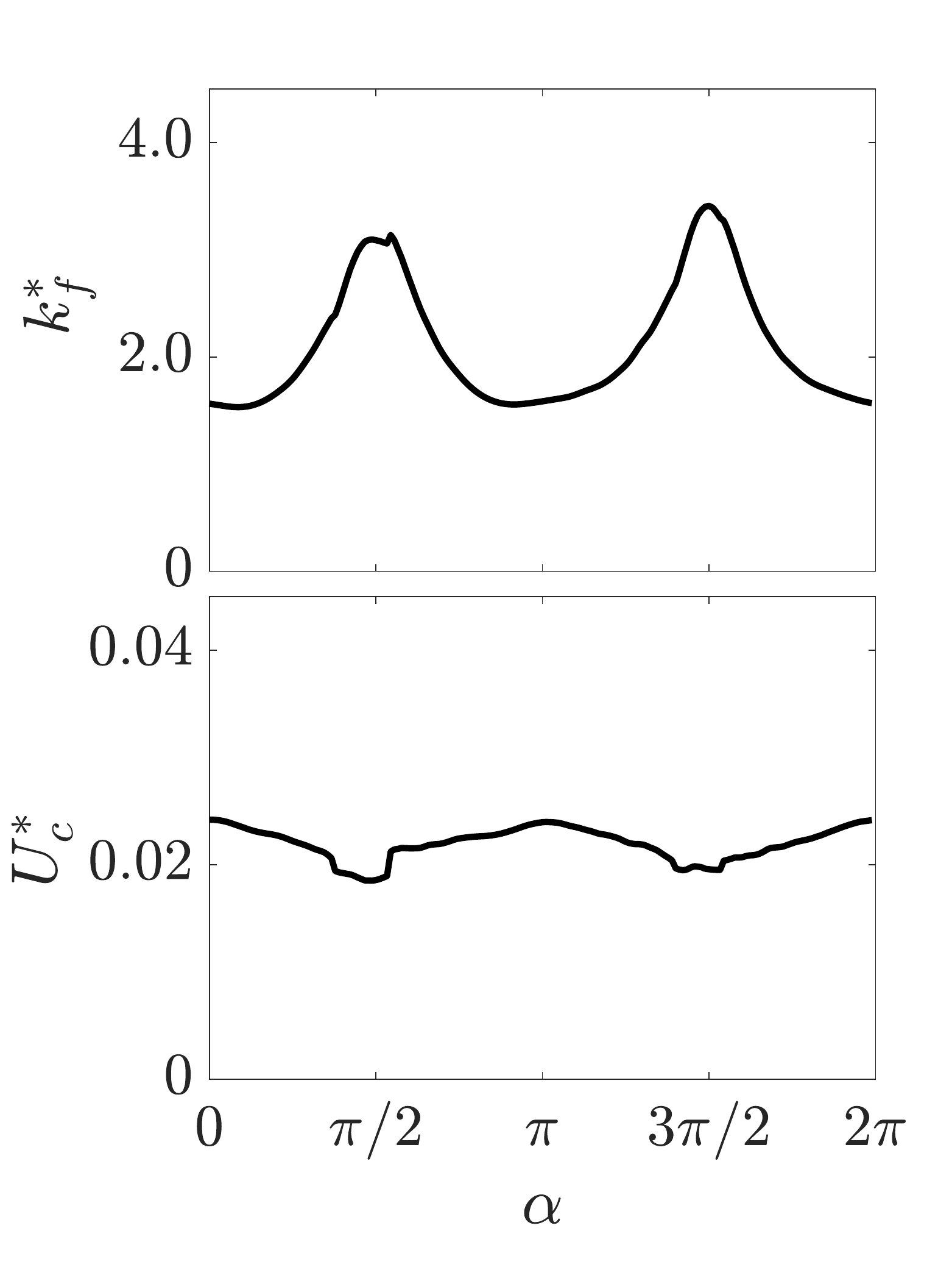}\label{fig:Res:CurveFitParam}}
   \caption[]{(a) Scaled front propagation as a function of scaled cylinder velocity $U_{c0}/U^*_{c}-1$. The curve fit with equation~\eqref{eq:front_norm_vel_curvfit} is indicated by the solid black curve, while the shaded area represents the rms from the curve fit. The error bar associated with each point is one standard deviation from the collection of time series used at a specific $\alpha$ and $U_{c0}$. The inset shows the front propagation compared to the cylinder velocity for $\alpha =0$ and $\alpha =\pi/2$. (b)~Resulting parameters for the curve fit in (a) as a function of $\alpha$.} 
   \label{fig:Res:NormFrontProp_Fit} 
\end{figure}

The inset in figure~\ref{fig:Res:NormalizedFrontPropagation} shows $k_f-U_{c0}$ curves for two angles $\alpha=0$ (red) and $\alpha=\pi/2$ (blue). $\alpha=0$ is an example of transverse front propagation, while $\alpha=\pi/2$ is representative of longitudinal front propagation. Here we also see that the difference between transverse and longitudinal front propagation is roughly a factor of 2~\cite{Han2016}. When normalizing the data with $k_f^*$ all data collapse on one curve as indicated in the main figure \ref{fig:Res:NormalizedFrontPropagation}. In figure \ref{fig:Res:CurveFitParam}, the two curve fit parameters $U^*_{c}$ and $k_f^*$ are plotted as functions of $\alpha$. Here, $k_f^*$ has two strong peaks at $\alpha=\pi/2$ and $\alpha=3\pi/2$. $U^*_{c}$, on the other hand does not show a strong dependence on $\alpha$.

\floatsetup[figure]{style=plain,subcapbesideposition=top}

\begin{figure}
    \sidesubfloat[]{\includegraphics[width=0.4\textwidth]{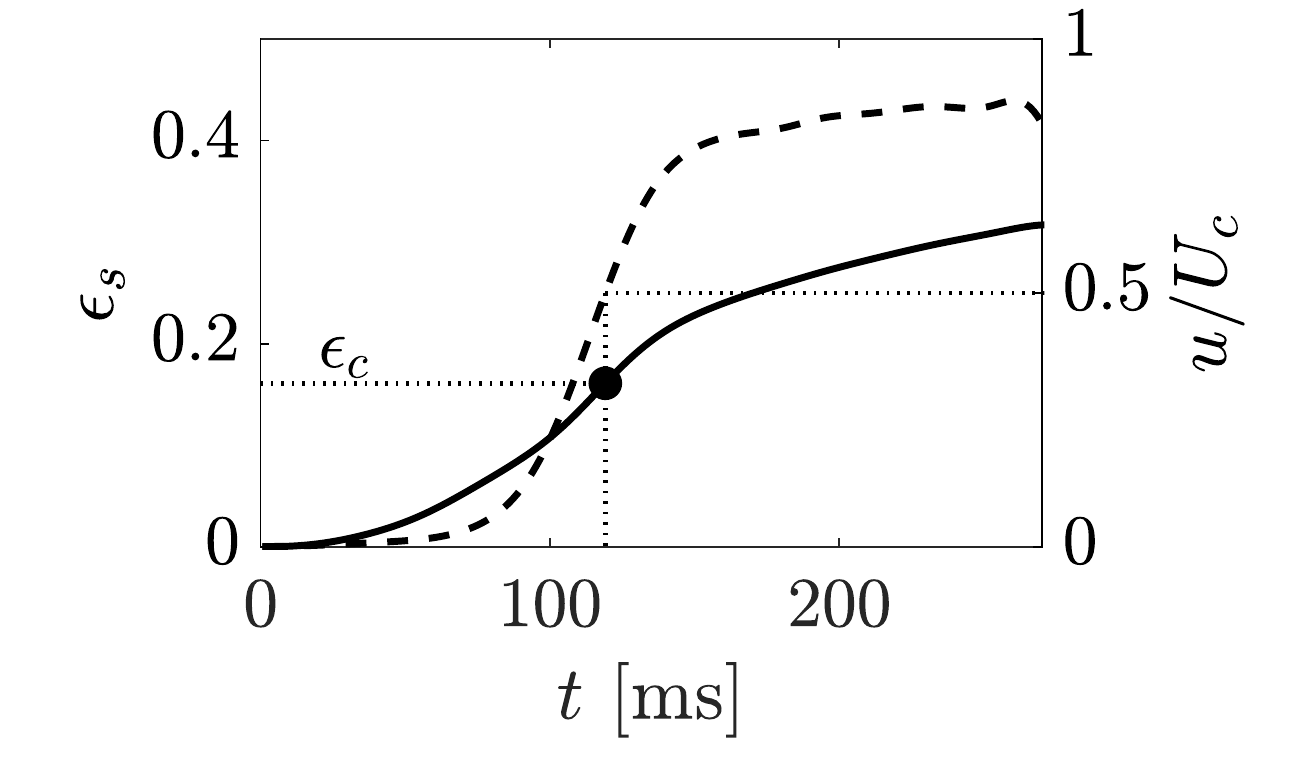}\label{fig:Res:strainhist}}
    \sidesubfloat[]{\includegraphics[width=0.4\textwidth]{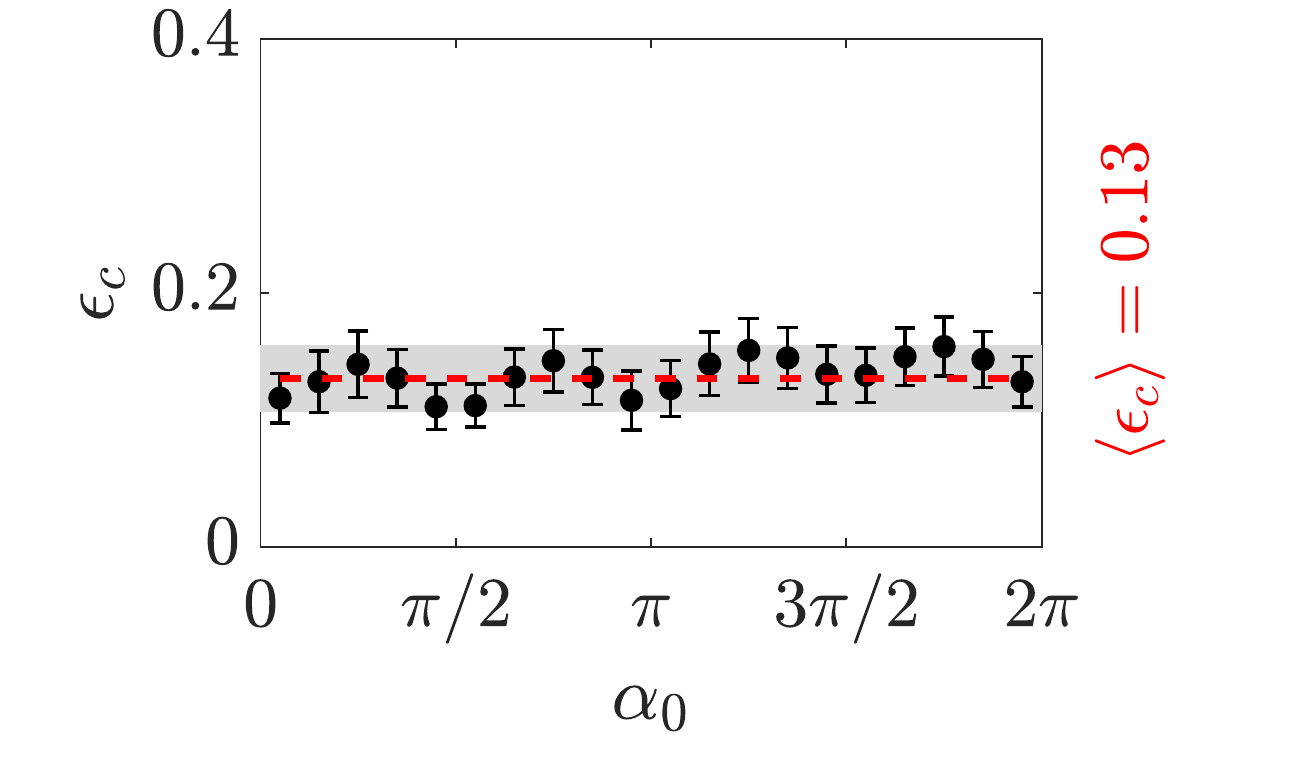}\label{fig:Res:straincrit}}
   \caption{Onset strain. (a) This plot shows the definition of $\epsilon_c$. The dashed line indicates velocity, while the solid line indicates accumulated strain. The dotted lines are simply for reference. This onset strain is defined as the strain at the point in time where the jamming front passes through that specific material point. The plot shows velocity and accumulated strain of a specific material point. As seen in the figure, $\epsilon_c=\epsilon_s|_{u=0.5U_{c}}$. (b)~Onset strain as a function of initial angle the material point has with the cylinder ($\alpha_0$). The data is taken from all cases where $U_{c0}/U_c^*-1>2.2$. The onset strain is approximately independent of position relative to the cylinder compared to the front propagation, which has a clear transverse-longitudinal anisotropy as seen in figure~\ref{fig:Res:FrontPropagationAlpha}. We do note that there is a weak longitudinal asymmetry in the accumulated strain. The onset strain is observed to be slightly higher behind the cylinder. This is also evident from the color band in figure \ref{fig:Res:StrainField}. }
   \label{fig:Rescritstraindef} 
\end{figure}

Even though the front travels with different velocities in the transverse and longitudinal directions, it is noted from figure \ref{fig:Res:StrainField} that the accumulated strain ($\epsilon_s$) has a roughly constant value in close proximity to the jamming front. That is, there is an onset strain associated with the transition from a liquid-like to solid-like state. We quantify this onset strain by limiting our analysis to experiments where $k_f$ is independent of $U_{c0}$. In our case, this is when $U_{c0}/U_c^*-1\ge2.2$. By looking at the strain history associated with a material point, we define the onset strain ($\epsilon_c$) as the point in time where $u=0.5U_{c}$. That is, when the jamming front passes through the material point. This can be seen in figure \ref{fig:Res:strainhist}. In order to check for dependence with the location of the material points, we look at the mean and spread of $\epsilon_c$ as a function of the angle the material points initially have with the cylinder ($\alpha_0$). We do this by bundling the data for the onset strain in $\sim 18^\circ$ intervals, and calculating the mean and standard deviation. This can be seen in figure \ref{fig:Res:straincrit} together with the global mean. As shown by \citet{Han2016}, the fact that the jamming front propagates with different speeds in the longitudinal and transverse directions is a direct consequence of the existence of an onset strain. By following the material points, we show that the accumulated strain at the jamming front has an approximately constant value for all angles about the cylinder. This shows that there is an underlying strain controlling the liquid-like to solid-like transition in the material, as is expected for systems of similar nature \cite{Han2016, Han2018, Wang2018, Pastore2011, Bi2011, Majmudar2005, Khandavalli2015, Fall2012}. We do note, however, that the onset strain shows slightly higher values behind the cylinder.

\floatsetup[figure]{style=plain,subcapbesideposition=top}

\begin{figure}
  \centering
  \includegraphics[width=0.85\textwidth]{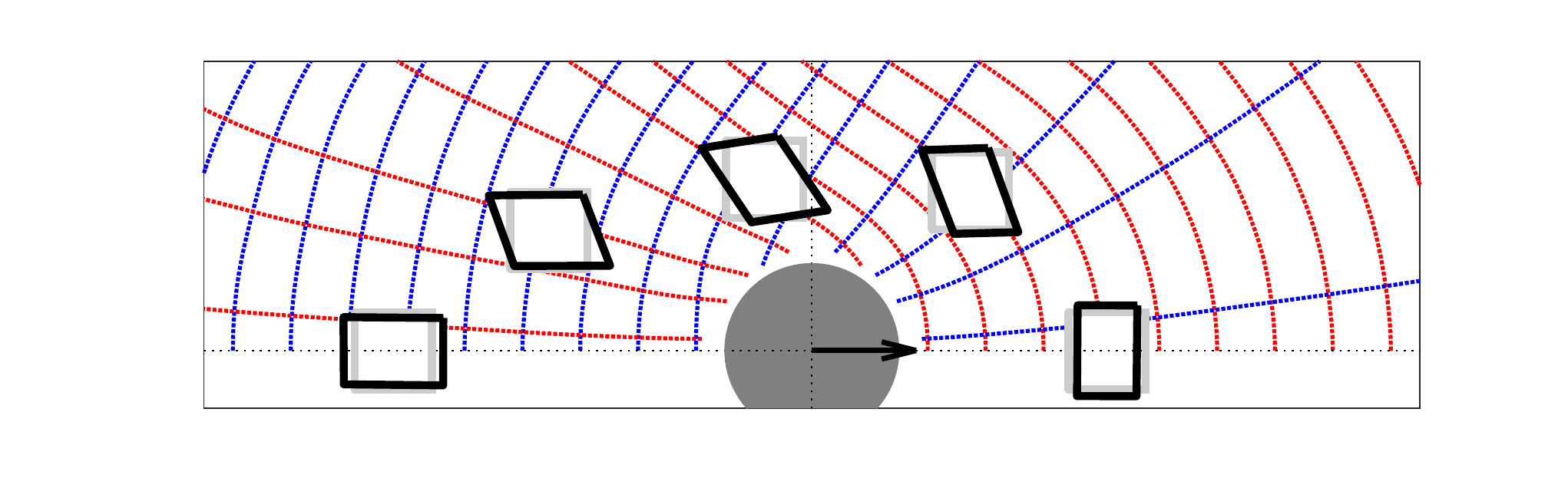}
  \caption{Schematic representation of the deformation around the cylinder. This schematic is drawn using real data from the $U_{c0}/U_c^* - 1 = 5.4$ case, at the end of the experiment where the region shown here has jammed. The scaled fluid parcels follow the flow. The dashed lines show the direction of principle elongation (red) and compression (blue). Note that the fluid parcel in the transverse position has rotated more than what one would expect from a simple shear. See \cite{SupMat} for a time resolved representation of the schematic.}
  \label{fig:Res:DefSchem}
\end{figure}

For the quasi 1D, transient, simple shear system described in \cite{Han2018, Han2019b}, a model linking the front propagation $k$ with the packing fraction $\phi$ was developed. In the experiment described here, we observe many of the phenomena expected from experiment of similar nature \cite{Peters2014, Waitukaitis2012, Han2016, Peters2016, Han2019}. That is, a constant front propagation with a relation $\backsim1:2$ between the transverse and longitudinal direction.  However, using the formula $1/k^*=-\gamma^*\ln\left(\Phi\right)$ \cite{Han2018, Han2019b} ($\Phi$ being a re-scaled packing fraction) yields a value of $k^*=6.6$ which is significantly different from our value of $k_f^*=1.6$ in the transverse direction. Neither the variance in our measured $k$, nor the uncertainty in $\phi$ explains this discrepancy. However, in our system the perturbation is both back and forth in the lab frame. Even though the suspension has come to rest before it is perturbed again, we expect the distribution of particles in the suspension to still retain some anisotropy due to the strain history from the previous perturbation. This anisotropy manifests itself as the numerical value $\gamma^*$ not being applicable in our case. It is also worth pointing out that our system is 2D, and we have no estimation for the full stress state in our system, as is the case in the quasi 1D, transient, simple shear system described in \cite{Han2018, Han2019b}. 2D effects can be seen in figure \ref{fig:Res:RotField}, showing the rotation and the movement of the material points. Notice from the deformed grid that the material in the transverse direction seems to have rotated more than one would expect from a simple shear in the lab frame. In the transverse direction it is primarily the velocity gradient component $\partial u / \partial y$ that dominates, but an extra rotation comes from the fact that material is being pushed out in front of the cylinder, and sucked into the wake. This creates a $\partial v/\partial x$ component with an opposite sign than the $\partial u / \partial y$ component. As we calculate strain rate as the sum $\sim\partial v/\partial x + \partial u / \partial y$ and rotation rate as $\sim\partial v/\partial x - \partial u  /\partial y$, in an accumulated sense we expect the material points to have rotated more than one would expect from a simple shear, as seen from the deformed grids in figure \ref{fig:Res:RotField} and the schematic in figure \ref{fig:Res:DefSchem}.

\begin{figure}
    \sidesubfloat[]{\includegraphics[trim = 0 100 0 0, clip, width=0.6\textwidth]{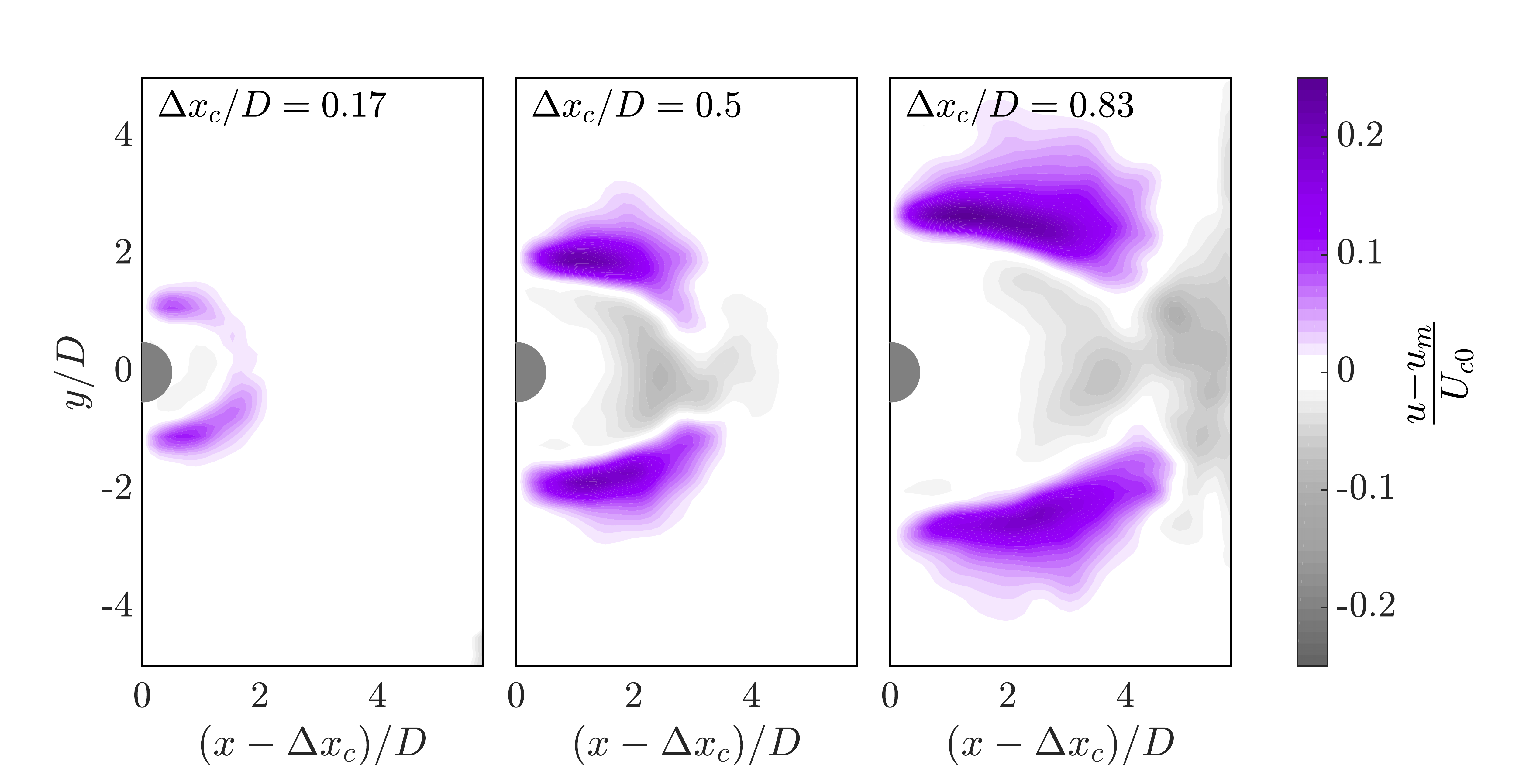}\label{fig:Res:VDiff}}\\
    \sidesubfloat[]{\includegraphics[trim = 0 100 0 0, clip,width=0.6\textwidth]{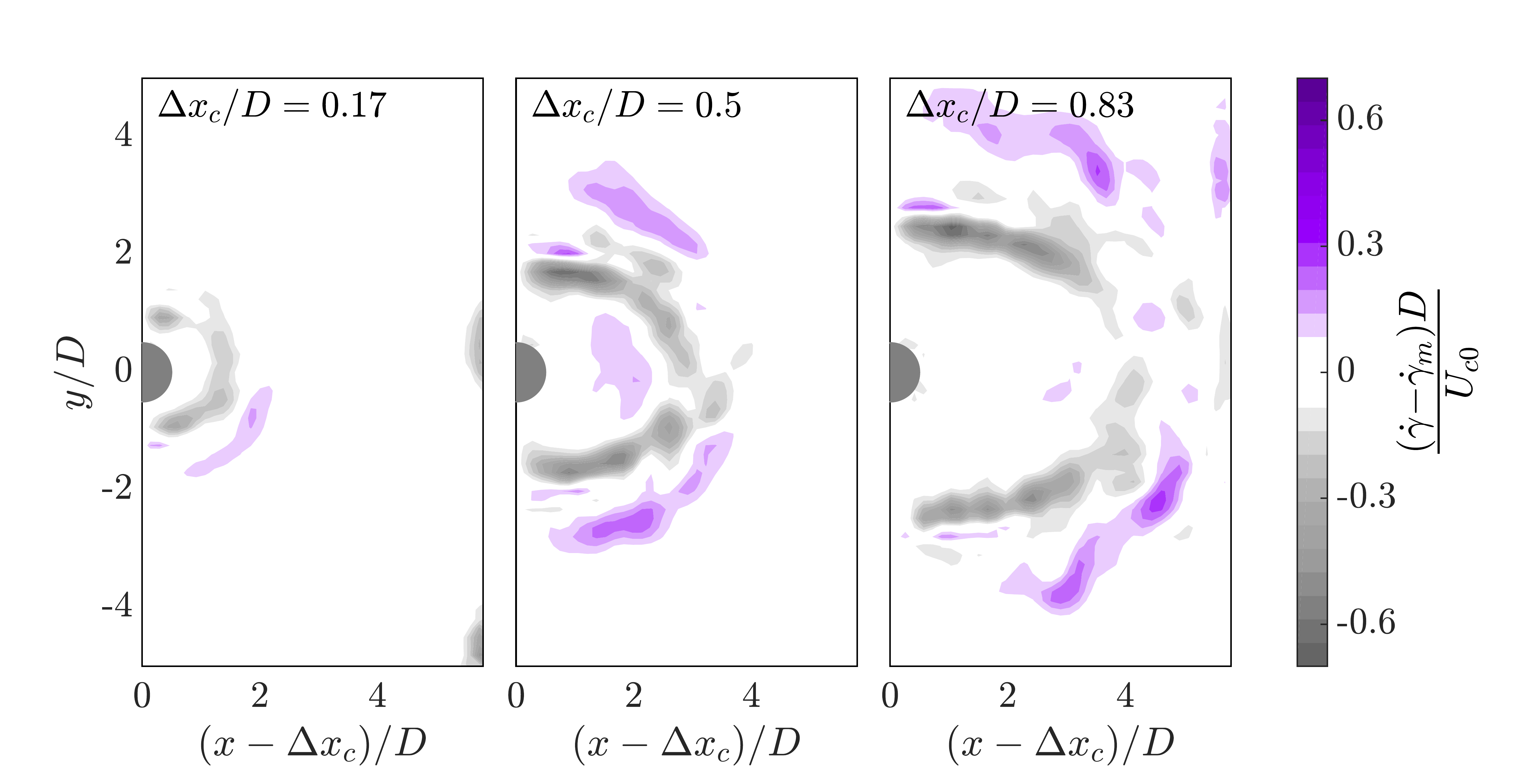}\label{fig:Res:SRDiff}}\\
    \sidesubfloat[]{\includegraphics[trim = 0 0 0 0, clip,width=0.6\textwidth]{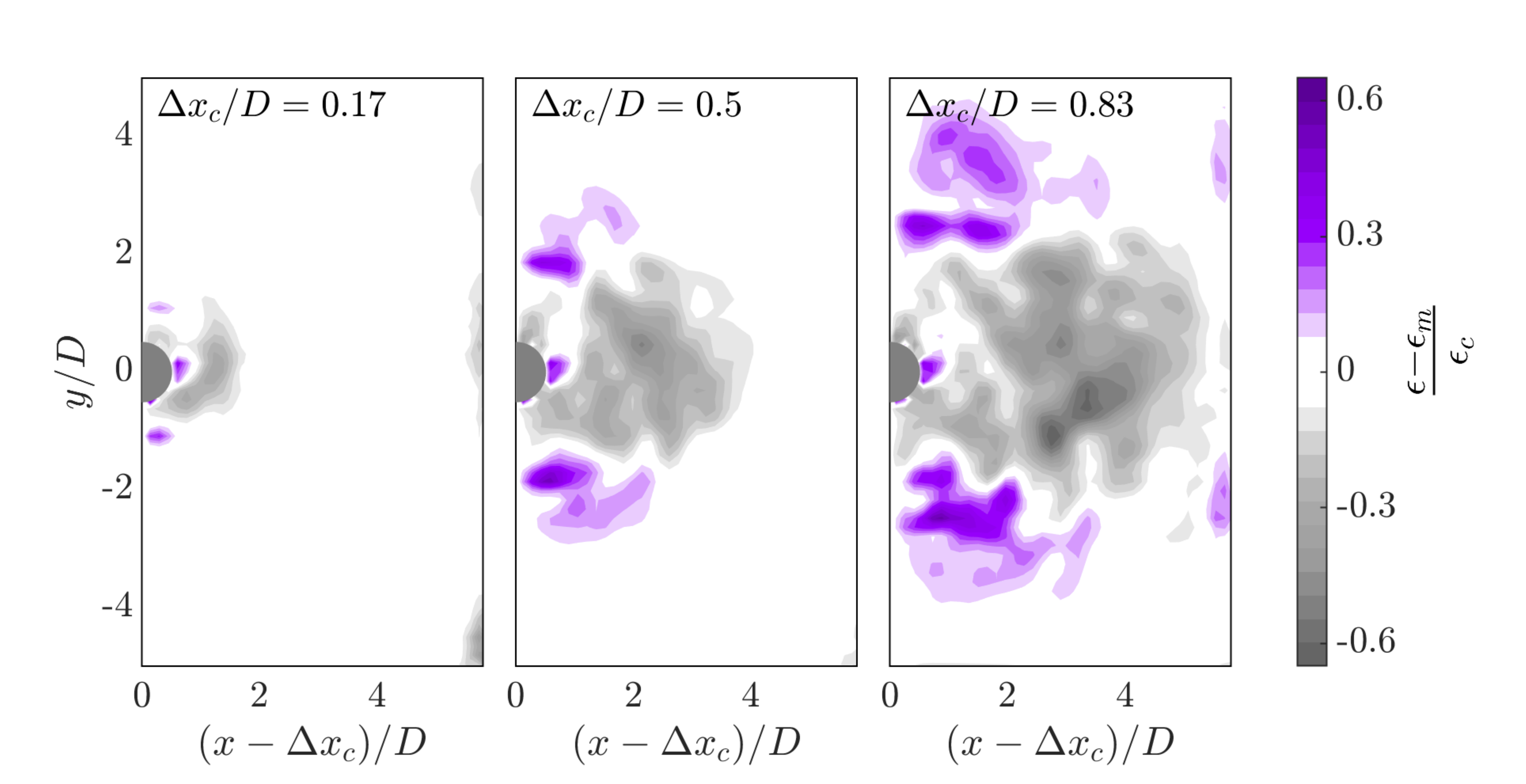}\label{fig:Res:SDiff}}
   \caption{Fore-aft asymmetry for the (a)~velocity, (b)~strain rate, and (c)~accumulated strain.  The asymmetry is computed by mirroring the fields in figure~\ref{fig:Res:Vel_rate_strain} about the centre of the cylinder and plotting the difference between the fore and aft fields. }
   \label{fig:Res:MirrDiff} 
\end{figure}

Although our results are symmetric in the transverse direction, a fore-aft asymmetry has been consistently observed throughout this work. We quantify this asymmetry by mirroring the velocity ($u$), strain rate ($\dot{\gamma}_s$), and strain ($\epsilon_s$) fields from figure~\ref{fig:Res:Vel_rate_strain} about the cylinder position and plotting the difference (figure \ref{fig:Res:MirrDiff}). The mirrored fields are denoted with the subscript $m$. While the fore-aft difference in the onset strain ($\epsilon_c$) and the front propagation factor ($k_f$) is roughly the same magnitude as the measurement variance, there are significant and consistent differences in the mirrored kinematic fields as illustrated in figure~\ref{fig:Res:MirrDiff}. Specifically, the peak differences between the fore and aft regions of the cylinder are approximately 20\%, 50\%, and 50\% for the velocity, strain rate, and accumulated strain, respectively. These differences are computed relative to the cylinder velocity, the strain rate defined by $U_{c0}/D$, and the onset strain, respectively. This indicates that while the magnitudes of $k_f$ and $\epsilon_c$ may only be mildly asymmetric, the spatial orientation of the fields is strongly asymmetric. This finding is novel and accessible for the first time in the present experimental configuration as previous set-ups could not simultaneously generate jamming fronts through pushing, pulling and shearing.

\floatsetup[figure]{style=plain,subcapbesideposition=top}


\section{Conclusion \& Discussion}

Experiments of a cylinder towed through a cornstarch-sucrose-water suspension were presented. The main goal was to investigate how the suspension behaved around a towed body in the domain where dynamic jamming occurs. By capturing high-speed images of the suspension surface, PIV was used to compute 2D, time-resolved velocity fields. We demonstrate novel methodology where the full field surrounding the cylinder is visualized and pulling, pushing and shearing are observed simultaneously for the first time. Here, we present measures of overall strain and rotation acquired from a polar decomposition of the deformation gradient tensor, inspired by common deformation measures in solid mechanics. This study focused on the region in time where the cylinder was moving with a constant speed, and the front propagated through the suspension unaffected by the boundary.

Though the setup is different from previous studies, we see several similarities. For a sufficiently high cylinder velocity, the dimensionless front propagation speed is independent of cylinder velocity, and the front travels roughly twice as fast in the longitudinal compared to the transverse direction. In addition, an onset strain of $\epsilon_c=0.13$ accompanies the jamming front as it propagates through the suspension regardless of position relative the cylinder. 

Even though our results are symmetrical in the  transverse direction, a longitudinal (fore-aft) asymmetry has been consistently observed throughout this work. Particularly, the velocity ($u$), strain rate ($\dot{\gamma}_s$), strain ($\epsilon_s$) and rotation ($\psi$) in figure~\ref{fig:Res:Vel_rate_strain} all indicate this. By mirroring the fields about the cylinder position, the asymmetry is quantified in figure~\ref{fig:Res:MirrDiff}. The fore-aft asymmetry is also visible in the front propagation factor $k_f$ and the onset strain $\epsilon_c$ (figure~\ref{fig:Res:straincrit}), to a lesser degree.

As noted in the discussion above, it has been shown how strain influences the flow for simpler systems \citep{Han2019b}, where strain reduces down to a scalar. Other experiments investigating shear jamming fronts in more complex systems focus only on the behaviour in front of \cite{Han2016,Han2019,Peters2014} or behind \cite{Majumdar2017} the perturbing body separately, thus the jamming front produced in these studies propagates in the fore or aft half-planes only. How the 2D flow develops around a translating body, producing jamming fronts in the full circumference simultaneously, has not been observed until now. Given the moderately large strains observed in our experiment (larger than a few percent \citep{Irgens2008}) there is no \emph{a priori} basis to assume fore-aft symmetry. With our system, we are able to observe and quantify a fore-aft asymmetry. To our knowledge, this is the first observation of this phenomenon in shear jamming dense suspensions.

\begin{acknowledgments}
RJH acknowledges the financial support of the Research Council of Norway (Grant No.~288046). IRP acknowledges financial support from the Royal Society (Grant No.~RG160089). Data supporting this study are openly available from the University of Southampton repository \cite{SupDat}.
\end{acknowledgments}


\appendix

\section{\label{app:FrontProp}Jamming front and front propagation}

From the PIV analysis we get a transient 2D velocity field $\mathbf{u}=\{u, v\}$. The jamming front is defined as the points in the suspension where the velocity is half the velocity of the cylinder \cite{Han2016, Waitukaitis2012, Peters2014, Peters2016, Majumdar2017}. The cylinder is moving in the positive $x$-direction with velocity  $U_c$, such that $\mathbf{U_c}=\{U_c, 0\}$. From figure \ref{fig:Exp:Def} we define the location of the jamming front relative to the cylinder as
\begin{equation}
\label{eq:front_pos}
    \mathbf{x_f} = \mathbf{x}|_{u=0.5U_c} - \mathbf{x_c}. 
\end{equation} 
This involves a Galilean transformation; that is, we look at the location of the jamming front relative to the cylinder. Furthermore, the velocity of the jamming front is defined as
\begin{subequations}
\begin{align}
\label{eq:front_norm_vel}
    u_{f} & = \left(\frac{d}{d t}\mathbf{x_f}\right) \cdot \mathbf{n_f} \\
                & = \left[\frac{d}{d t}\left(\mathbf{x}|_{u=0.5U_c} - \mathbf{x_c}\right)\right] \cdot \mathbf{n_f} \\
                & = \left(\frac{d \mathbf{x}|_{u=0.5U_c}}{d t} - \mathbf{U_c}\right) \cdot \mathbf{n_f},
\end{align}
\end{subequations}
where the jamming front normal vector $\mathbf{n_f}$ is calculated as
\begin{equation}
\label{eq:front_norm_vec}
    \mathbf{n_f} = \begin{pmatrix}0 & 1\\-1 & 0 \end{pmatrix} \frac{d\mathbf{x_f}}{d\alpha}\frac{1}{|\frac{d\mathbf{x_f}}{d\alpha}|}. 
\end{equation}
As the jamming front does not correspond to material points, but a velocity contour, we define its propagation direction to be in the direction of the jamming front normal vector $\mathbf{n_f}$. Figure~\ref{fig:Exp:VelProf} shows the connection between the velocity profile and the location of the jamming front.

Front propagation factor ($k_f$) is defined in equation~\ref{eq:front_norm_vel_norm}. We will focus on averaged results. For a given $\alpha$ and $U_c$, the front propagation $k_f$ is presented as an average over the time window of analysis for all relevant cases (See figure~\ref{fig:Exp:FrontPosTime}). This is calculated from a minimum of 13 cases for each set of $\alpha$ and $U_c$.

In regions where the front normal vector is pointing in the transverse ($\mathbf{n_{ft}}=\{ 0, 1\}$) or longitudinal ($\mathbf{n_{fl}}=\{1, 0\}$) direction, the relations
\begin{subequations}
\begin{align}
\label{eq:front_norm_vel_translong}
 k_{ft} &= \frac{1}{U_c}\frac{d}{d t}(y|_{u=0.5U_c}), \\
 k_{fl} &= \frac{1}{U_c}\frac{d}{d t} (x|_{u=0.5U_c}) - 1  
\end{align}
\end{subequations}
are recovered, similar to \citet{Han2016}.

\section{\label{app:Strain}Strain, strain rate and rotation}
Data for strain and strain rate are presented as scalar fields in the main text. We define the scalar strain rate as the shear rate magnitude \cite{Irgens2008}
\begin{equation}
    \label{eq:strainrate_measure}
    \dot{\gamma}_{s}=\sqrt{2||\mathbf{D}||^2},
\end{equation}
and the strain as
\begin{equation}
    \label{eq:strain_measure}
    \epsilon_s=\sqrt{||\mathbf{e}||^2}.
\end{equation}
Here, $\mathbf{D}=\frac{1}{2}(\frac{\partial \mathbf{u}}{\partial \mathbf{x}}+\frac{\partial \mathbf{u}}{\partial \mathbf{x}}^T )$ represents the strain rate tensor, and $\mathbf{e}$ is the Eulerian logarithmic strain measure. Logarithmic strain is defined based on the strain increment $d\epsilon=dl/l$ and is calculated as \cite{Irgens2008}
\begin{equation}
    \label{eq:logstrain}
    \epsilon = \int\frac{dl}{l}=\ln\left(\frac{l_f}{l_0}\right)=\ln(\lambda).
\end{equation}
Here, $l_f$ and $l_0$ are the final and initial lengths of a line segment, while $\lambda$ is a principle stretch. This definition of strain is in some sources referred to as Hencky-, natural- or true strain \cite{Bazant1998, Fitzgerald1980, Rees2006, Irgens2008}. Here, we use the definition of logarithmic strain from \cite{Nasser2004}, which has the spectral representation
\begin{equation}
    \label{eq:strain_measure_tensor}
    \mathbf{e}=\sum_i \ln(\lambda_i) \mathbf{n}_i\otimes\mathbf{n}_i.
\end{equation}
$\lambda_i$ and $\mathbf{n}_i$ are the principle values and principle directions of the left stretch tensor $\mathbf{V}$. Note that $\mathbf{V}$ is a positive definite symmetric tensor with positive real principle values. The following steps are taken when calculating the left stretch tensor, and ultimately $\epsilon_s$:
\begin{enumerate}
    \item In order to calculate the stretch in the material, an estimate of the movement of the material points is needed. $\mathbf{X}=\{X,Y\}$ refers to a material point in our initial configuration at $t=0$. At any time $t$, a material point, located at $\mathbf{X}$ in the initial configuration, has moved to the point $\mathbf{x_p}(\mathbf{X},t)=\{x_p(t), y_p(t)\}$. The movement of a material point is estimated from the PIV data as \cite{Boutelier2019} 
    \begin{subequations}
    \begin{align}
        \label{eq:mat_point}
        x_p(t_n) & = X +  \sum_{t_i=0}^{t_n}u[x_p(t_i),y_p(t_i), t_i)]\Delta t\\
        y_p(t_n) & = Y +  \sum_{t_i=0}^{t_n}v[x_p(t_i),y_p(t_i), t_i)]\Delta t,
    \end{align}
    \end{subequations}
    where $\Delta t$ is given by the framerate.
    \item  With the movement of the material points, we are able to calculate the deformation gradient tensor ($\mathbf{F}=\frac{\partial \mathbf{x_p}}{\partial \mathbf{X}}$). The left stretch tensor $\mathbf{V}$ is acquired by a polar decomposition $\mathbf{F}=\mathbf{V}\mathbf{R}=\mathbf{R}\mathbf{U}$ \cite{Irgens2008, Rees2006, Nasser2004}. 
    \item The principle values ($\lambda_i$) and principle directions ($\mathbf{n}_i$) of $\mathbf{V}$ are calculated. We order the principle values such that $\lambda_1>\lambda_2$, which results in $\lambda_1$ and $\mathbf{n}_1$ signify the amount and direction of extension, while $\lambda_2$ and $\mathbf{n}_2$ signify the amount and direction of compression. By closer inspection of equations \eqref{eq:strain_measure} and \eqref{eq:strain_measure_tensor}, we calculate the strain as $\epsilon_s=\sqrt{(\ln(\lambda_1))^2+(\ln(\lambda_2))^2}$.
\end{enumerate}

We choose the left stretch tensor $\mathbf{V}$ rather than the right stretch tensor $\mathbf{U}$, as the strain will be presented in the deformed configuration. This has no consequence for equation \eqref{eq:strain_measure}, as the principal values of $\mathbf{U}$ and $\mathbf{V}$ are equal. It is the principal directions that differ. The principal directions of $\mathbf{U}$ signifies the axis of principal stretch relative to the initial configuration, while the principal directions of $\mathbf{V}$ signifies the axis of principal stretch relative to the deformed configuration.

We also present the rotation of the material points $\psi$. This angle is calculated from the rotation matrix obtained from the polar decomposition above. As seen from the formulas $\mathbf{U}=\mathbf{R}^T\mathbf{V}\mathbf{R}$ and $\mathbf{R}^T\mathbf{R}=\mathbf{I}$, $\mathbf{R}$ represents a change of basis between the principle stretches viewed in the reference configuration and the deformed configuration. In 2D, the elements of $\mathbf{R}$ are
\begin{equation}
    R = \begin{pmatrix}\cos(\psi) & -\sin(\psi)\\\sin(\psi) & \cos(\psi) \end{pmatrix},
\end{equation}
and the rotation angle of the material points is simply calculated as $\cos(\psi)=\frac{1}{2} \text{tr} \,\mathbf{R}$.





\providecommand{\noopsort}[1]{}\providecommand{\singleletter}[1]{#1}%

\end{document}